\documentclass[aps,prl,twocolumn,showpacs,amsmath,amssymb,floats]{revtex4}
\usepackage{graphicx}
\usepackage{dcolumn}
\begin{document}

\def\Journal#1#2#3#4{{#1} {\bf #2}, #3 (#4)}
\def\NCA{\rm Nuovo Cimento}
\def\NIM{\rm Nucl. Instrum. Methods}
\def\NIMA{{\rm Nucl. Instrum. Methods} A}
\def\NPB{{\rm Nucl. Phys.} B}
\def\PLB{{\rm Phys. Lett.}  B}
\def\PRL{\rm Phys. Rev. Lett.}
\def\PRD{{\rm Phys. Rev.} D}
\def\PRC{{\rm Phys. Rev.} C}
\def\ZPC{{\rm Z. Phys.} C}
\def\JPG{{\rm J. Phys.} G}
\def\st{\scriptstyle}
\def\sst{\scriptscriptstyle}
\def\mco{\multicolumn}
\def\epp{\epsilon^{\prime}}
\def\vep{\varepsilon}
\def\ra{\rightarrow}
\def\ppg{\pi^+\pi^-\gamma}
\def\vp{{\bf p}}
\def\ko{K0}
\def\kb{\bar{K0}}
\def\al{\alpha}
\def\ab{\bar{\alpha}}
\def\be{\begin{equation}}
\def\ee{\end{equation}}
\def\bea{\begin{eqnarray}}
\def\eea{\end{eqnarray}}
\def\CPbar{\hbox{{\rm CP}\hskip-1.80em{/}}}
%
\title{\large \bf A Study of the Quasi-elastic (e,e'p) Reaction on $^{12}$C, 
$^{56}$Fe and $^{97}$Au.}
\author {
D.~Dutta$^{15,10,a}$\footnotetext{\noindent $^a$ Present address:Duke University, Durham, NC 27708},
D.~van Westrum$^{4,b}$\footnotetext{\noindent $^b$ Present address:Micro-g Solutions Inc., Erie, CO 80516},
D.~Abbott$^{22}$,
A.~Ahmidouch$^{7}$,
Ts.~A.~Amatuoni$^{25}$,
C.~Armstrong$^{24,c}$\footnotetext{\noindent $^c$ Present address:TJNAF, Newport News, VA 23606},
J.~Arrington$^{2,d}$\footnotetext{\noindent $^d$ Present address:Argonne Nat'l Lab, Argonne, IL 60439},
K.~A.~Assamagan$^{6,e}$\footnotetext{\noindent $^e$ Present address:Brookhaven Nat'l Lab, Upton, NY 11973},
K. Bailey$^1$,
O.~K.~Baker$^{22,6}$,
S.~Barrow$^{18}$,
K.~Beard$^6$,
D.~Beatty$^{18,c}$,
S.~Beedoe$^{14}$,
E.~Beise$^{11}$,
E.~Belz$^{4}$,
C.~Bochna$^{9}$,     
P.~E.~Bosted$^{12}$,
H.~Breuer$^{11}$,
E.~E.~W.~Bruins$^{10}$,
R.~Carlini$^{22}$,
J.~Cha$^{6,f}$\footnotetext{\noindent $^g$ Present address:Mississippi State University, Mississippi State, MS 39762},
N.~Chant$^{11}$,
C.~Cothran$^{23}$,
W.~J.~Cummings$^{1}$,
S.~Danagoulian$^{14}$,
D.~Day$^{23}$,
D.~DeSchepper$^{10}$,
J.-E.~Ducret$^{21}$,
F.~Duncan$^{11,g}$\footnotetext{\noindent $^f$ Present address:Queens University, Kingston, Ontario, Canada},
J.~Dunne$^{22,f}$,
T.~Eden$^6$,
R.~Ent$^{22}$,
H.T.~Fortune$^{18}$,
V.~Frolov$^{19,h}$\footnotetext{\noindent $^h$ Present address:University of Minnesota, Minneapolis, MN 55439},
D.~F.~Geesaman$^1$,
H.~Gao$^{9,10,a}$,
R.~Gilman$^{22,20}$,
P.~Gu\`eye$^6$,
J.~O.~Hansen$^{1,c}$,
W.~Hinton$^{6,c}$,
R.~J.~Holt$^{9}$,
C.~Jackson$^{14}$,
H.~E.~Jackson$^1$,
C.~Jones$^{1}$,
S.~Kaufman$^1$,
J.~J.~Kelly$^{11}$,
C.~Keppel$^{22,6}$,
M.~Khandaker$^{13}$,
W.~Kim$^8$,
E.~Kinney$^4$,
A.~Klein$^{17}$,
D.~Koltenuk$^{18,i}$\footnotetext{\noindent $^i$ Present address:Lincoln Labs, MIT, Lexington, MA 02420},
L.~Kramer$^{10,j}$\footnotetext{\noindent $^j$ Present address:Florida International University, Miami, FL 33199},
W.~Lorenzon$^{18,k}$\footnotetext{\noindent $^k$ Present address:University of Michigan, Ann Arbor, MI 48109},
A.~Lung$^{2,c}$,
K.~McFarlane$^{13,l}$\footnotetext{\noindent $^l$ Present address:Hampton University, Hampton, VA 23668},
D.~J.~Mack$^{22}$,
R.~Madey$^{6,7}$,
P.~Markowitz$^5$,
J.~Martin$^{10}$,
A.~Mateos$^{10}$,
D.~Meekins$^{22,c}$,
M.~A.~Miller$^{9}$,
R.~Milner$^{10}$,
J.~Mitchell$^{22}$,
R.~Mohring$^{11}$,
H.~Mkrtchyan$^{25}$,
A.~M.~Nathan$^{9}$,
G.~Niculescu$^{6,m}$\footnotetext{\noindent $^m$ Present address:University of Virginia, Charlottesville, VA 22901},
I.~Niculescu$^{6,n}$\footnotetext{\noindent $^n$ Present address:James Madison University, Harrisonburg, VA 22807},
T.~G.~O'Neill$^1$,
D.~Potterveld$^1$,
J.~W.~Price$^{19,o}$\footnotetext{\noindent $^o$ Present address:University of California, Los Angeles, CA 90095},
J.~Reinhold$^{1,j}$,
C.~Salgado$^{13}$,
J.~P.~Schiffer$^1$,
R.~E.~Segel$^{15}$,
P.~Stoler$^{19}$,
R.~Suleiman$^{7,p}$\footnotetext{\noindent $^p$ Present address: MIT, Cambridge, MA 02139},
V.~Tadevosyan$^{25}$,
L.~Tang$^{22,6}$,
B.~Terburg$^{9,q}$\footnotetext{\noindent $^q$ Present address: General Electric Corp., Cleveland, OH 44114},
Pat Welch$^{16}$,
C.~Williamson$^{10}$,
S.~Wood$^{22}$, 
C.~Yan$^{22}$,
Jae-Choon~Yang$^{22}$
J.~Yu$^{18}$,
B.~Zeidman$^1$,
W.~Zhao$^{10}$,
and B.~Zihlmann$^{23}$.
}

\affiliation{
$^1${\underbar{Argonne National Laboratory, Argonne IL 60439}} \\
$^2${\underbar{California Institute of Technology, Pasadena CA  91125}} \\
$^3${\underbar{Chungnam National University, Taejon 305-764 Korea}} \\
$^4${\underbar{University of Colorado, Boulder CO 80309}} \\
$^5${\underbar{Florida International University, University Park, FL 33199}} \\
$^6${\underbar{Hampton University, Hampton VA 23668}}\\
$^7${\underbar{Kent State University, Kent OH 44242}} \\
$^8${\underbar{Kyungpook National University, Taegu, South Korea}} \\
$^{9}${\underbar{University of Illinois, Champaign-Urbana IL 61801}} \\
$^{10}${\underbar{Massachusetts Institute of Technology, Cambridge MA 02139}}\\
$^{11}${\underbar{University of Maryland, College Park MD 20742}} \\
$^{12}${\underbar{University of Massachusetts, Amherst MA 01003}}\\
$^{13}${\underbar{Norfolk State University, Norfolk VA 23504}}\\
$^{14}${\underbar{North Carolina A \& T, Greensboro NC 27411}}\\  
$^{15}${\underbar{Northwestern University, Evanston IL 60201}}\\
$^{16}${\underbar{Oregon State University, Corvallis OR 97331}}\\
$^{17}${\underbar{Old Dominion University, Norfolk, VA 23529}}\\
$^{18}${\underbar{University of Pennsylvania, Philadelphia PA 19104}}\\
$^{19}${\underbar{Rensselaer Polytechnic Institute, Troy NY 12180}}\\
$^{20}${\underbar{Rutgers University, New Brunswick NJ 08903}}\\
$^{21}${\underbar{CE Saclay, Gif-sur-Yvette France}} \\
$^{22}${\underbar{Thomas Jefferson National Accelerator Facility,
Newport News VA 23606}} \\
$^{23}${\underbar{University of Virginia, Charlottesville VA 22901}}\\
$^{24}${\underbar{William and Mary, Williamsburg, VA 23187}}\\
$^{25}${\underbar{Yerevan Physics Institute, Yerevan, Armenia}}\\
}
\begin{abstract}
We report the results from a systematic study of the quasi-elastic (e,e'p) 
reaction on $^{12}$C, $^{56}$Fe and $^{197}$Au performed at Jefferson Lab. We
 have measured nuclear transparency and extracted spectral functions (corrected for 
radiation) over a Q$^2$ range of 0.64 - 3.25 (GeV/c)$^2$ for all three nuclei. 
In addition we have extracted separated longitudinal and transverse spectral 
functions at Q$^2$ of 0.64 and 1.8 (GeV/c)$^2$ for these three 
nuclei (except for  $^{197}$Au at the higher Q$^2$). The spectral functions 
are compared to a number of theoretical calculations. The measured spectral 
functions differ in detail but not in overall shape from most of the theoretical 
models. In all three targets the measured spectral functions show considerable 
excess transverse strength at Q$^2$ = 0.64 (GeV/c)$^2$, which is much reduced 
at 1.8 (GeV/c)$^2$.
\end{abstract}
\maketitle

\section*{INTRODUCTION}
The value of studying electronuclear reactions has long been recognized.  In 
such studies the entire nucleus is accessed via a well-understood interaction.  
A new avenue of investigations has been opened up with the completion of the 
continuous beam, multi-GeV electron accelerator at the Thomas Jefferson National 
Accelerator Facility, also known as Jefferson Lab (JLab). The present paper 
reports results from the first experiment done at this facility, which is a study 
of (e,e'p) reactions in the quasi-elastic region.  This experiment utilized one 
of the advantages of electron scattering, namely, the transferred energy and 
momentum can be varied separately, and one of the main features of JLab, namely, 
the high intensity continuous electron beam of CEBAF which makes it possible to 
do coincidence measurements orders of magnitude more extensive than could be 
done previously.  

The  simplest  model of  a  nucleus  is  one of  independent  nucleons
populating  the  lowest available  shell-model  orbits.  In a simple
picture of $e-p$ scattering within a  nucleus, the electron scatters from a
single protons which is moving due to its Fermi momentum. The struck proton 
may then interact with the residual A-1 nucleons before leaving the nucleus.  
Of course,  neither the nucleus nor the scattering process  are this simple 
and the deviations from  these  simple  pictures  reveal  much  about  
nuclei  and  their constituents, both real and virtual.  The present 
experiment consisted
of  measuring   proton  spectra  in   coincidence  with  inelastically
scattered electrons with the energy of the electrons chosen such as to
be in  the "quasi-elastic" region,  i.e. at energies  corresponding to
scattering  from  single off-mass-shell  nucleons.   The spectra  were
taken in an angular region about the "conjugate" angle, i.e. the angle
for  scattering  from  stationary  nucleons,  over  an  angular  range
sufficient to cover the smearing  of the two-body kinematics caused by
the Fermi momentum of the  confined protons.  Data were taken over the
range 0.64~$<$~Q$^2~<$~3.25~(GeV/c)$^2$  where Q$^2$ is  the square of the 
four-momentum transferred to the struck proton.

For an electron knocking a proton, $p$, out of a nucleus $A$ with energy 
transfer $\omega$ and (three) momentum transfer $\vec{q}$ leaving a scattered 
proton, $p'$, and a residual nucleus, $A - 1$, two important kinematic 
quantities are the missing energy:
\begin{equation}
E_m = \omega - T_{p'} - T_{A-1}
\end{equation}
\noindent
and missing momentum:
\begin{equation}
\vec{p}_m = \vec{p}_{p'} - \vec{q}
\end{equation} 
\noindent
where $T_{p'}$ and $T_{A-1}$ are the kinetic energy of the knocked out proton 
and recoiling nucleus, respectively.  The 
spectral functions were extracted from the $E_m$ and $\vec{p}_m$ spectra and compared to a variety of theoretical calculations.  The total (e,e'p) yields are 
obtained by integrating over the spectral functions and the transparencies 
then determined by comparing these yields with those predicted by Plane Wave 
Impulse Approximation (PWIA) calculations.  Because the PWIA does not allow for final-state interactions the ratio of measured to calculated yield should just be the 
fraction of outgoing protons which do not suffer a final-state interaction and 
this is what is defined to be the transparency.  Determinations of nuclear 
transparencies using the (e,e'p) reaction have been reported for a range of 
targets covering the periodic table, at Bates for Q$^2$ = 0.34 (GeV/c)$^2$ 
~\cite{gerry92}, at SLAC for Q$^2$ between 1 and 7 (GeV/c)$^2$~\cite{ton95, mak94}, and 
more recently at JLAB between 3 and 8.1 (GeV/c)$^2$~\cite{garrow01}.  The 
present work maps out regions not previously covered and is of greater 
statistical accuracy.  Longitudinal - Transverse (L - T) separations were 
performed at two values of Q$^2$ from which the first reported extensive 
separated spectral functions are obtained.  Some transparency results from the 
present experiment have been previously published ~\cite{prl1}, as have the 
separated spectral functions for carbon~\cite{prc2}.

The differential cross section for elastic electron-proton scattering is given 
by the well-known Rosenbluth formula:
\begin{equation}
\label{cc1}
\frac{d\sigma}{d\Omega} = \left(\frac{d\sigma}{d\Omega}\right)_
{\mbox{\small Mott}}\frac{Q^2}{|\vec{q}|^2}[G_{E}^{2}(Q^2) + 
\tau\epsilon^{-1}G_{M}^{2}(Q^2)]
\end{equation}
\noindent                                                   
where $\left(\frac{d\sigma}{d\Omega}\right)_{\mbox{\tiny Mott}}$ is the differential 
cross 
section for the scattering of an electron off a unit point charge, $\epsilon =\frac{1}{1+2(1+\tau)\tan^2(\frac{\theta}{2})}$ is the 
virtual polarization parameter, ${\tau = {{|\vec{q}|^2}\over{Q^2}} - 1}$,  G$_E$ 
is the proton electric form factor and G$_M$ is 
the proton magnetic form factor in units of the nuclear magneton,  
$\frac{e\hbar}{2M_{p}c}$ where $M_p$ is the proton mass.

The L - T separation is performed by measuring the cross section at different 
values of $\epsilon$ while keeping Q$^2$ constant, thus permitting the 
extraction of G$_E$  and  G$_M$.

In scattering from a nucleus the cross section is expressed in terms of 
four response functions and in the PWIA the coincidence 
(e,e'p) cross sections can be written:
$$
\frac{d^6\sigma}{dE_{e'}d\Omega_{e'}dE_{p'}d\Omega_{p'}}= p'E_{p'}\sigma_
{\mbox{\tiny Mott}}$$
$$
[\lambda^{2}W_{L}(q,\omega) + [\frac{\lambda}{2}+ \tan^{2}(\frac{\theta}{2})]W_{
T}(q,\omega)+
$$
\begin{equation}
\label{cc2}
\lambda[\lambda + \tan^{2}(\frac{\theta}{2})]^{1/2}W_{LT}(q,\omega)\cos(\phi) + 
\frac{\lambda}{2}W_{TT}(q,\omega)\cos(2\phi)].
\end{equation}
 
\noindent
where $\lambda =Q^2/|\vec{q}|^2$, $\theta$  is the scattering angle and $\phi$  
is the azimuthal angle between the scattering plane and the  plane containing 
$\vec{q}$ and $\vec{p'}$.

The physics of interest is contained in the 4 response functions W$_L$, W$_T$, 
W$_{LT}$ and W$_{TT}$.  Both of the interference terms, W$_{LT}$ and W$_{TT}$ 
are proportional to sin$\gamma$, where $\gamma$  is the angle between the 
scattered proton and the transferred momentum $\vec{q}$.  Therefore, when 
measurements are made along $\vec{q}$, i.e. in "parallel kinematics", the 
interference terms are absent.  Varying the incident energy makes it possible to 
vary $\theta$ at constant $q$ and $\omega$ and thus disentangle W$_L$ and W$_T$, 
that is, perform an L - T separation.  Although, the position of the 
spectrometers allowed measurements only in the scattering plane, the 
interference term W$_{LT}$ could be investigated by varying the proton angle 
about the direction of  $\vec{q}$. Measurements were taken by varying both 
$\theta$ and $\gamma$.  This is the first L - T separation measured for quasi-elastic (e,e'p) scattering that covers a large 
range in both A and Q$^2$.

\section*{EXPERIMENT}

\subsection*{Electron Beam}
The experiment was performed in 1995~-~1996 in Hall C at JLab and, was the first experiment performed at the Laboratory.  Data were taken at (nominal) 
electron energies E$_e$ = (0.8N + 0.045) GeV with N = 1 - 4 representing the 
number of "passes" the electrons made around the accelerating track.  The 
absolute beam energy was determined at one-pass by two independent methods.  One 
method is to use the inelastic scattering to an excited state whose energy is 
accurately known to calibrate the dispersion of a spectrometer and then use 
the calibrated spectrometer to measure the energy of the scattered electron as a function of nuclear target mass.  For 
these measurements a carbon target was used and the dispersion determined by 
measuring the difference in position of the electrons scattered to the ground 
and the 4.43891 $\pm$ 0.00031 MeV~\cite{offerman91} first excited state.  A BeO 
target was then substituted and the energy of the beam, E, determined using the 
formula:

\begin{equation}
\Delta E_{\mbox{\small recoil}} = 2E^2\sin{\frac{\theta}{2}}^2(\frac{1}{M_1}-
\frac{1}{M_2})
\label{disp}
\end{equation}

One can accurately determine $\Delta E_{\mbox{\small recoil}}$  because once the 
dispersion 
has been accurately measured the only unknown in Eq.~\ref{disp} is the beam 
energy $E$.  This procedure was repeated for several values of the spectrometer magnetic field.  With both 
targets a small correction was made for the energy loss of the electrons in the 
target.

The other method is to determine the angle of the diffraction minimum for 
scattering to a state where the position can be accurately calculated.  The 
minimum for scattering to the $^{12}$C first excited state has been determined 
to be at $Q^2$ = 0.129$\pm$0.0006 (GeV/c)$^2$~\cite{fas85}.  The (four) momentum 
transfer can be written:                                         

\begin{equation}
Q^2=4EE'\sin^2{\frac{\theta}{2}}, E'= \frac{E}{1 + 
\frac{2E\sin^2{\frac{\theta}{2}}}{M}}
\label{qande}
\end{equation}

\noindent
where $M$ is the mass of the scattering nucleus and $\theta$ is the electron scattering angle.  An improvement in accuracy in 
the measurement of $Q$ is obtained by using the ratio of elastic scattering to 
inelastic scattering.  Again, then, the only unknown is the incident electron 
energy $E$.  The two methods agreed to 1 part in 2000 and the absolute energy 
determination using these methods is believed to be accurate to 10$^{-3}$.  
These methods become less feasible as the energy is increased.  The beam energy 
can also be determined by measuring the energy and angle of the scattered 
particles in electron-proton elastic scattering.  Because of the uncertainties 
in the angle and momentum measurements this method is less accurate than the 
other two but has the advantage that it can be used over the entire range of 
incident electron energies.  Elastic $e-p$ scattering was used to measure the 
energy of the three-pass beam with an uncertainty of 1 part in 500.  Beam energy 
was also determined by measuring the magnetic field needed to bend the beam around 
the Hall C arc. The energy calibration as well as other 
aspects of the experiment are discussed more completely elsewhere~\cite{thesis}.

Beam currents of 10 to 60 $\mu$Amps were used.  The currents were monitored by 3 
microwave cavities that were installed for this purpose in the Hall C beam 
line\cite{jlab1}.  The absolute calibration was performed by comparison with 
an Unser cavity, which is a parametric DC current transformer with very stable 
gain but a drifting offset which was determined as part of our daily calibration procedure.  The overall accuracy in the 
beam current measurement was $\pm$1\%.

\subsection*{Targets} 

Data were taken with  $\approx$ 200 mg/cm$^2$ C, Fe and Au targets mounted on 
a steel ladder in an aluminum scattering chamber. The target thicknesses were 
determined to about 0.1\%.  The $e-p$ elastic scattering data used for 
calibration were taken using the 4.0~cm cell of the Hall C cryogenic target~\cite{jlab2}. During the early part of the experiment, before the cryogenic target 
was available, some data were taken with a solid CH$_2$ target but these data 
were used to check some kinematic offsets only.  The compositions of hydrocarbon targets are subject to change under beam irradiation and therefore all the 
calibration data were taken with the liquid hydrogen target.  The cryogenic 
targets are also mounted on a ladder with both ladders contained in the 
aluminum scattering chamber.  The 123.0~cm diameter scattering chamber has entrance 
and exit snouts for the beam and several pumping and viewing ports. The 
particles that went to the High Momentum Spectrometer (HMS) spectrometer 
exited through a 0.4~mm aluminum window and those to the Short Orbit Spectrometer (SOS) through a 0.2~mm aluminum window.  For both spectrometers the 
particles had to pass through about 15~cm of air before entering the 
spectrometer.  

\subsection*{Spectrometers} 

Data were taken with the HMS and the SOS in coincidence.  This experiment served 
as the commissioning experiment for these spectrometers.  The HMS detected the 
electrons and the SOS the protons, except at the highest Q$^2$ where the roles 
of the spectrometers were reversed.

\subsubsection*{High Momentum Spectrometer}

The HMS is a 25$^o$ vertical bend spectrometer made up of superconducting 
magnets in a QQQD configuration.  The dipole field is monitored and regulated 
with an NMR probe and kept constant at the 10$^{-4}$ level.  The spectrometer 
rotates on a 
pair of rails between 12.5$^o$ and 90$^o$ with respect to the beam line.  The 
HMS maximum central momentum is 7.3 GeV/c and in preparing for the present 
experiment the spectrometer was tested up to 4.4 GeV/c although the highest 
setting at which data were taken was 2.6 GeV/c.  The usable momentum bite is 
of the spectrometer is $\approx$20\%. A momentum resolution ($\sigma$) of $<$1.4~10$^{-3}$, 
and an in-plane (out-of-plane) angular resolution of 0.8~(1.0)~mrad was 
achieved for the HMS.  With no collimator in place the solid angle subtended for a
point target is 8.1~msr.  A  6.35~cm thick HEAVYMET (machinable Tungsten alloy,10\% Cu Ni; 
density~=~17 g/cm$^3$) collimator with a flared octagonal aperture 
limited the solid angle to 6.8~msr.  The 
higher momentum 
particles were usually detected in the HMS and except at the backward (electron) 
angles these were the electrons.  Detailed information about the HMS can be 
found in~\cite{jlab3}.

\subsubsection*{Short Orbit Spectrometer}

The SOS consists of 3 (normal conducting) magnets in a QD$\overline{\mbox{D}}$ 
configuration.  The deflection is vertical with the net bend of 18$^o$ at the 
central momentum.  The magnetic fields are monitored with Hall probes. With its 
short path length of 11 m this spectrometer is particularly well suited for 
detecting short-lived particles though obviously this attribute was not used in 
the present experiment.  The spectrometer can be moved between 13.1$^o$ and 
168.4$^o$ with respect to the beam line (during this experiment the minimum angle was 14.5 $^o$) and can be moved up to 20$^o$ out of the 
horizontal plane, though this was not done in this experiment.  The 
spectrometer maximum central momentum is 1.8 GeV/c with a nominal 
momentum bite of 40\%. A momentum resolution ($\sigma$) of $<$~1.0~10$^{-3}$, 
an inplane (out-of-plane) angular resolution of 4.5(0.5) msr was achieved for 
the SOS. The solid angle subtended is  $\approx$ 9~msr for a point target, although a collimator similar 
to that used with the HMS limited the solid angle to 7.5~msr.  As with the HMS, 
further details about the SOS can be found 
in the spectrometer documentation~\cite{jlab3,moh3}.

\subsubsection*{Detector Stacks}

The detector stacks in the two spectrometers are virtually identical.  The 
particles pass through, in order, a set of drift chambers, a pair of hodoscopes, 
a gas \v{C}erenkov detector, another pair of hodoscopes and then a lead-glass 
calorimeter.  The particle velocity is inferred 
from the time-of-flight between the two pairs of hodoscopes though the spectra 
proved to be so clean that it was not necessary to use time-of-flight for 
particle identification.  Signals from the hodoscope planes provide the 
trigger and in the electron arm particle identification can be incorporated 
into the trigger by requiring a signal from the 
\v{C}erenkov counter and/or a sufficiently large pulse from the calorimeter.  
Coincidences between the triggers selected out the (e,e'p) events that make up 
the physics data.

The drift chambers serve to determine the particles' position, x (y), and 
direction, x' (y'), in the bend (nonbend) plane of the spectrometer and it is 
these quantities that are used to reconstruct the events.  Each spectrometer 
has two chambers and each chamber contains 
six planes of wires. In each HMS chamber one pair measures x, one pair measures y 
and the remaining two planes 
are rotated $\pm$15$^o$ with respect to the x plane.   
The purpose of the third pair of planes is to correlate the xy information when 
more than one particle traverses a 
chamber during the readout interval.  In the SOS chambers one pair is in the x plane 
and the other two pairs of planes are at $\pm$60$^o$ with respect to the x plane.
Position resolution per plane is 
$<$~250~$\mu$m in the HMS chambers and $<$~200~$\mu$m in the SOS.  The wire 
chamber data was used to reconstruct the trajectory of the particles and 
determine the particles momentum fraction relative to the central momentum, 
$\delta$p/p.

Wire chamber tracking efficiency is an important element in the overall system 
efficiency and, as such, must be accurately measured.  This was done by using 
the position information in the hodoscopes to tag particles passing through a 
small central region of the chambers and then see what fraction of such events 
was reconstructed from the wire chamber signals.  In both spectrometers 
typical tracking efficiency was greater than 97\% which was determined to 
better than 1\%.  The main sources of wire chamber tracking inefficiency are 
inefficiencies in the chambers themselves (we require 5 of the 6 planes have good hits) and inefficiency in the reconstruction algorithm. The measured 
inefficiency was the sum of these inefficiencies and no attempt was 
made to disentangle the two.

\subsection*{Calibrations}

\subsubsection*{Spectrometer optimizations}

Because this was the first experiment performed in Hall C, considerable effort 
went into first optimizing the performance of the spectrometers and then optimizing the data analysis so as to achieve the highest possible accuracy.  The magnetic field of the HMS quadrupoles was mapped to determine its optical axis and its effective field length versus current, with effective field length defined as the line integral of the field divided by the average field. However, the HMS 
dipole was not mapped and its magnetic field to current (B-to-I) relation was 
calculated using the TOSCA program~\cite{tosca}. The measured field map 
of the quadrupole and the TOSCA generated map of the dipole were used to build an optics model of the spectrometer with the COSY program~\cite{cosy1}. For a 
desired magnetic field of the dipole (i.e. a desired central momentum) the dipole current was set according to the B-to-I relation predicted by the TOSCA program, while the COSY model was used to get the starting value of the quadrupole to dipole ratio (Q/D). The Q/D ratio was then varied to get the best focus in the 
spectrometer and these optimized ratios were used to determine the 
current settings of the quadrupole for a desired central momentum of the 
spectrometer. From elastic $e-p$ scattering data it was later determined that 
the B-to-I relation of the dipole predicted by TOSCA was wrong by about 0.9\%. 
The dipole currents were adjusted accordingly to correct for this difference.  
A similar procedure was followed for the SOS except that the quadrupole was 
not mapped and the optics model was formulated using the COSY program assuming 
the field of the quadrupole magnet to be an ideal quadrupole.  The SOS 
dipole B-to-I relation was also found to be slightly wrong (0.55\%) and suitable corrections were made to the setting procedure.

The basic strategy in determining the momentum and direction of the scattered 
particles is to use the wire chamber data to determine the position, (x,y), and 
the angles, (x',y'), of the particles at the focal plane which, in turn, 
specifies the trajectory of the particle through the spectrometer.  This of 
course requires knowing the fields of the spectrometer, which are represented by 
a set of matrix elements that relate the position and direction of the particles 
as they cross the focal plane, to the particle's momentum, angles of emission, 
and starting position along the beam direction.  The accuracy of the final 
results then depends on how well the matrix elements simulate the spectrometers 
and hence a great deal of effort went into optimizing these matrix elements.
 
The COSY program was used to calculate an initial set of 
reconstruction matrix elements using the mapped fields for the HMS magnets and 
the SOS dipoles and an assumed pure quadrupole field for the SOS quadrupole. 
The Hall C Matrix Element Optimization Package CMOP~\cite{cosy2} was used to 
optimize the reconstruction matrix elements. In this package 
the dispersion matrix elements are optimized using 
momentum scans, i.e. varying the central momentum by varying the magnetic fields. 
For each spectrometer these momentum scans were performed for both elastic p(e,e') 
and elastic $^{12}$C(e,e') scattering.  
In order to obtain the angular matrix elements sieve slits, which are 
collimators containing accurately positioned holes, were placed in front of 
each of the spectrometers so that rays of known initial position and direction 
could be traced. The angular matrix elements were then fit by the CMOP package 
(using singular value decomposition method) to accurately reproduce the known 
positions of the sieve slit holes. Similarly the target $y$ position 
(projection of the target length along the beam) reconstruction was 
optimized by utilizing the CMOP package with data from scans along the 
beam direction. These scans were performed by raising and lowering a 
slanted carbon target and the continuum portion of the carbon spectrum was 
used. Most of these calibration data were taken at 
one-pass, 845 MeV, with a check for reproducibility made with two-pass, 1645 MeV, 
electrons. 

\subsubsection*{Acceptances}

The spectrometer's acceptances were studied with the aid of the 
simulation code SIMC, which is an adoption to the JLab Hall C 
spectrometers of the (e,e'p) simulation code written for SLAC experiment 
NE18~\cite{makth}. 
This simulation package employs models for each of 
the spectrometers (HMS and SOS).  The same models were also used to study the 
optical properties of the spectrometers.  These models use COSY generated 
sets of matrices to simulate the transport of charged 
particles through the magnetic field of the spectrometer to each major aperture 
of the spectrometer.  Energy loss and multiple scattering in the intervening 
material were also included.  The events that passed through all apertures were 
then reconstructed back to the target using another set of matrices generated by 
COSY.  Surviving events were assigned a weight based on the PWIA cross-section, 
radiative corrections and coulomb corrections.  The PWIA cross-section was 
calculated using the deForest~\cite{defor83} prescription $\sigma_{cc1}$ for the 
off-shell $e-p$ cross-section and an Independent Particle Shell Model (IPSM) 
spectral function for the target nucleus involved.  The PWIA calculations and 
the IPSM spectral functions are elaborated in the next two sections.  The 
radiative corrections in SIMC were performed according to the 
Mo and Tsai~\cite{motsai69} formulation adapted for the coincidence (e,e'p) 
reaction as described in Ref.~\cite{ent02}.  Further, a normalization factor 
was calculated from the experimental luminosity, phase space volume and 
the total number of events generated, so that the simulation provided a 
prediction of the absolute yield.

The reconstructed momentum, scattering angle, out-of-plane angle and target 
length distributions generated by the model were compared with the distributions 
obtained from the $e-p$ elastic scattering data as shown in Fig. 1.  These results 
are an indicator of how well the model acceptance simulated the true 
acceptance of the spectrometer. This was the status of the model during the experiment, there has been significant improvement in the model since then. 
\begin{figure}[htbp]
\begin{center}
\includegraphics*[width=9.5cm,height=10.0cm]{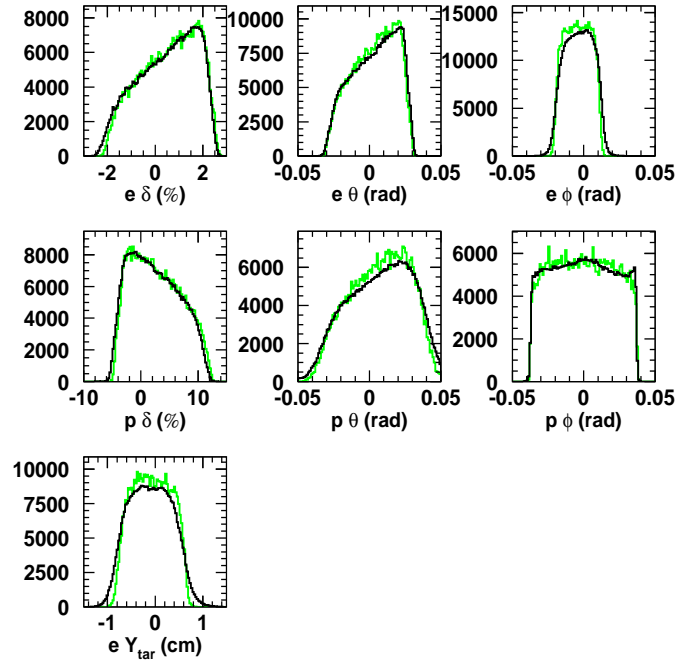}
\end{center}
\caption{Comparison of calculated (dark line) and measured (light line) 
distributions.  Top row is momentum, angle, and out of plane angle for electrons 
and the middle row the same for protons.  Last picture is the projection of 
the distribution along the target for electrons.} 
\label{fig1}
\end{figure}

\subsection*{Corrections} 
 
\subsubsection*{Radiative corrections} 

A major issue in electron scattering experiments is radiative corrections.  
The incoming and outgoing electrons can interact with the Coulomb field of the 
nucleus involved in the scattering which results in the emission and absorption 
of virtual photons and emission of real, primarily soft, photons.  Formulas for 
correcting for these radiative losses have been worked out by Mo and 
Tsai~\cite{motsai69}.  Correcting spectral functions deduced from (e,e'p) 
coincidence spectra is considerably more complicated because in this case the 
radiated momentum as well as the lost energy must be allowed for.  Although 
these are real physical processes they are experiment specific and so most 
theoretical calculations do not take them into account. The prescription for 
doing this for coincidence (e,e'p) reactions developed by Ent et al.~\cite{ent02} was used in the present work.   Using this prescription, 
radiated spectra are generated which can be directly compared with the 
experimentally measured spectra.  This point is discussed further in the 
section on spectral functions. 

\subsubsection*{Nuclear reactions}

Protons, being hadrons, will undergo strong interactions in traversing the 
detector stack and valid coincidences will be lost.  This loss was measured 
directly using $e-p$ elastic scattering.  Each scattered electron must 
have an accompanying proton and electrons were selected from a small region at 
the center of the acceptance thus insuring that protons could only be lost 
through nuclear interactions and other spectrometer inefficiencies.  
Transmissions of close to 95\% were measured for both spectrometers and are 
believed to be known to 1\%.  The absorption is virtually constant over the 
range of proton energies encountered in this experiment and therefore the small 
uncertainty in the absorption has little effect on any of the results. 

\subsubsection*{Deadtimes}
 
There were two data acquisition deadtimes of possible concern: electronic 
deadtime and computer deadtime.  Electronic deadtime occurs when triggers are 
not counted because the electronics hardware is busy processing previous 
triggers.  Electronic deadtime is dependent on the width of the logic signals, 
which for nearly all of the gates was 30 ns.  This deadtime was measured by 
recording the rates of multiple copies of the trigger with varying widths and 
then extrapolating to the rate at zero width. 
For both spectrometers the electronic deadtime was found to be $<$ 0.1\%.  
Computer deadtime is a more serious matter.  Most of the earlier data were taken 
in non-buffered mode where the processing time was
about 400~$\mu$s.  Later data were taken in the buffered mode with processing 
times of about 75~$\mu$s.  Over 80\% of the data were taken with deadtimes of 
$<$10\% but there were a few runs where deadtimes were as great as 60\%.  Even 
in these extreme cases the loss of event is known to better than 0.5\% from 
the ratio of the number of triggers generated to the number of triggers recorded by the data acquisition. This method was checked by measuring a large rate run
and then varying the fraction of triggers recorded by the data acquisition.

\section*{RESULTS}

\subsection*{Kinematics}
Table I shows the kinematics settings where data were taken.  The protons in the 
nucleus have finite momentum and therefore the struck protons from quasi-elastic 
scattering will emerge in a cone about the three-momentum transfer $\vec{q}$  
and measurements must be taken across this cone.  The lower the magnitude of 
$\vec{q}$ the broader the cone but, fortunately the cross section increases 
with decreasing Q$^2$.  While it is desirable to take data over as large a range 
of Q$^2$ as possible the cross section falls off so rapidly with increasing 
Q$^2$ that at the highest Q$^2$ point, 3.25 (GeV/c)$^2$, the cross section is so 
small that data could only be taken on one side of the conjugate angle.  L - T 
separations were performed at Q$^2$ of 0.64(GeV/c)$^2$ and 1.8(GeV/c)$^2$.   In 
order to get a good separation, data should be taken at as divergent values of 
$\epsilon$ (Eq 1) as possible, which translates into a large $\epsilon$ point at 
small (electron) angle and large incident energy and a low $\epsilon$ point at 
large angle and small energy (Table I).  The cross section decreases rapidly 
with increasing angle and so it was only possible to cover one side of the 
proton cone at $\epsilon$ = 0.31, Q$^2$ = 1.8(GeV/c)$^2$ and even at Q$^2$ = 
0.64(GeV/c)$^2$ there was time for only one point on the low-angle side of the 
cone.  Furthermore, no gold data were taken at the larger angle and higher 
Q$^2$ (1.8 GeV/c$^2$).

\begin{table}
\caption[Table of kinematics for Experiment E91-013]{Table of kinematics for 
Experiment E91-013, the central proton angles in bold represents the conjugate
angle.} 
\label{kmat}
\begin{center}
\begin{tabular}{|c|c|c|c|l|c|c|} \hline \hline
       &Central  &Central  & Central& Central   &          &        \\ 
Beam   &electron &electron & proton & proton    &Q$^2$     &        \\
Energy &Energy &Angle    & Energy & Angle     &          &$\epsilon$\\
(GeV)  &(GeV)  &(deg)   &  (MeV) & (deg)          &$\frac{GeV^2}{c^2}$ & \\ \hline 
\hline
       &         &         &        &           &          &         \\
       &         &         &        & 36.4,39.4 &          &         \\
       &         &         &        & 43.4,47.4 &          &         \\
 2.445 & 2.075   & 20.5    &  350   &51.4,\bf{55.4}& 0.64     & 0.93    \\
       &         &         &        & 59.4,63.4 &          &         \\
       &         &         &        & 67.4,71.4 &          &         \\
       &         &         &        & 75.4      &          &         \\ \hline
       &         &         &        &           &          &         \\
       &         &         &        & 27.8      &          &         \\
       &         &         &        & \bf{31.8} & 0.64     & 0.38    \\
 0.845 & 0.475   & 78.5    &  350   & 35.8,39.8,&          &         \\
       &         &         &        & 43.8,47.8 &          &         \\ \hline
       &         &         &        &           &          &         \\
       &         &         &        & 32.6.36.6,&          &         \\
 3.245 & 2.255   & 28.6    &  970   & \bf{40.6},& 1.80     & 0.83    \\
       &         &         &        & 44.6,48.6,&          &         \\
       &         &         &        & 52.6      &          &         \\ \hline
       &         &         &        &           &          &         \\
       &         &         &        & \bf{22.8},&          &         \\
 1.645 & 0.675   & 80.0    &  970   & 26.8,30.8 & 1.83     & 0.31    \\
       &         &         &        & 34.8      &          &         \\ \hline 
       &         &         &        &           &          &         \\
 2.445 & 1.725   & 32.0    &  700   & 31.5,35.5 & 1.28     &         \\
       &         &         &        & 39.5,\bf{43.5}&      & 0.81    \\
       &         &         &        & 47.5,51.4 &          &         \\
       &         &         &        & 55.4      &          &         \\ \hline
       &         &         &        &           &          &         \\
 3.245 & 1.40    & 50.0    & 1800   & \bf{25.5} & 3.25     & 0.54    \\
       &         &         &        & 28.0,30.5 &          &         \\ \hline 
\hline
\end{tabular}
\end{center}
\end{table}

\subsection*{Spectral Functions}

The spectral function for protons in a nucleus S$(E_s, \bf{p}_m)$ is defined as 
the probability of finding a proton with separation energy $E_s$ and momentum 
$\bf{p}_m$ inside that nucleus.  Obtaining spectral functions was a major 
objective of the present work and this section details how the spectral 
functions were deduced from the measured missing energy and missing momentum 
spectra.

\subsubsection*{Hydrogen Data}

\begin{figure}[htbp]
\begin{center}
\includegraphics*[width=9.5cm,height=13.0cm]{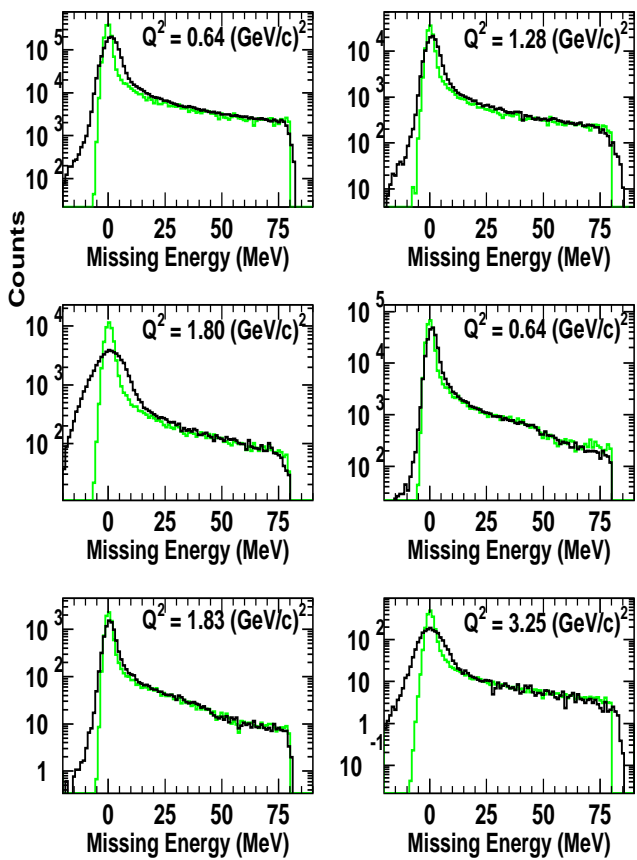}
\end{center}
\caption{Measured missing energy spectra for hydrogen (dark line) compared to 
spectra calculated using the Monte Carlo code SIMC(light line). The spectra with the same Q$^2$ refer to the forward and backward electron angle kinematics respectively for the L/T separation kinematics.}
\label{radtest}
\end{figure}

A missing energy and a missing momentum spectrum was obtained at each data point.  
For the hydrogen target this served as a measure of the response of the system 
while for the other targets these are the spectra from which the spectral 
functions are determined.  Hydrogen missing energy spectra along with the Monte 
Carlo calculated spectra at the various kinematics are shown in 
Fig.~\ref{radtest}.  The fact that the low energy tail is well reproduced out to 
the highest missing energy accepted (80 MeV), shows that the radiative 
corrections are being handled correctly.  Energy resolution, which is not of primary
importance in the present work, is clearly not well incorporated into the code 
in that the calculated zero missing energy peak is always narrower than 
that observed.  The peaks get broader with increasing energy of the 
scattered particle (see Table I), as could be expected, and this 
effect is not adequately accounted 
for.  The effect is most dramatic at the two values of Q$^2$ where data was 
taken at two different electron angles, and the peak is much broader at the 
forward angle where the 
electron energy is higher, while the proton energy remains the same.

The ratio of the observed to predicted $e-p$ elastic scattering yield is shown 
in Table II.  In calculating the predicted yield the electric form factor G$_E$ 
is taken to have the dipole form:
\begin{equation}
G_{E} = \left( 1 + \frac{Q^2}{0.71} \right)^{-2}
\end{equation}                                                                                                                             
and $G_M$ is taken from the Gari-Kr\"{u}mpelmann~\cite{gk73} parameterization 
which, to a good approximation, yields $G_M = \mu_{p}G_E$.  Rosenbluth 
separation measurements of $e-p$ scattering~\cite{walker94} support the validity 
of this relationship.

\begin{table}
\caption{Ratio of observed to predicted yield for $e-p$ elastic scattering.  
Uncertainties are statistical only, except for the (e,e'p) point at 3.25 
(GeV/c)$^2$ where there is an additional systematic uncertainty that is discussed 
in the text.} 
\label{hydrorat}
\begin{center}
\begin{tabular}{|c|c|c|c|} \hline \hline
        Q$^2$  & $\epsilon$ & \multicolumn{2}{c}{data/simulation}  \vline\\ \hline
    (GeV/c)$^2$  &            &   H(e,e'p) &   H(e,e')  \\ \hline                     
0.64      &  0.93      &   1.006 $\pm$ 0.005 & 1.015 $\pm$ 0.005 \\
     0.64      &  0.38      &   0.986 $\pm$ 0.005 & 0.997 $\pm$ 0.005 \\
     1.28      &  0.81      &   1.007 $\pm$ 0.005 & 1.009 $\pm$ 0.005 \\
     1.80      &  0.83      &   0.991 $\pm$ 0.005 & 1.003 $\pm$ 0.005 \\
     1.83      &  0.31      &   0.987 $\pm$ 0.005 & 0.989 $\pm$ 0.005 \\
     3.25      &  0.54      &   0.94  $\pm$ 0.012 $\pm$ 0.06 & 0.991 $\pm$0.007 
\\ \hline \hline
\end{tabular}
\end{center}
\end{table}

The typical systematic uncertainty for these measurements was 2.3\%. However, the
 large uncertainty in the (e,e'p) yield at Q$^2$ = 3.25 (GeV/c)$^2$ is due to 
an uncertainty in the proton efficiency due to malfunctioning wire chambers 
in the HMS.  For all of the other points, including the single-arm 
electrons at 3.25 (GeV/c)$^2$, calculated and measured yield agree to within 
about 1\%.  The setting for Q$^2$ = 3.25 (GeV/c)$^2$  was the only one at which 
the protons were detected in the HMS and this efficiency problem was corrected 
before the data on the complex nuclei was taken.
 
As an alternative to performing a Rosenbluth separation, a polarization transfer 
method has been developed~\cite{acg81} for measuring the ratio of the electric 
to the magnetic form factor and a recent experiment using this method reports 
that for the free proton  $\mu_{p}G_E/G_M$ decreases with increasing Q$^2$ 
declining to a value of 0.61 at Q$^2$ = 3.47 (GeV/c)$^2$~\cite{mjones00}.  A 
value of 0.79 is found at Q$^2$ = 1.8 (GeV/c)$^2$ while at Q$^2$ = 0.64 
(GeV/c)$^2$ it is only 5\% less than the
Q$^2$ = 0 value of unity.  In calculating the simulation cross sections for 
Table I the dipole (Eqn. 7) and Gari-Kr\"{u}mpelmann~\cite{gk73} values for G$_E$ 
and G$_M$, respectively, are used.  The implications of the results of Jones et. 
al. \cite{mjones00} for the present work are discussed in the section on L-T 
separations.  
 
\subsubsection*{Missing Energy Spectra for the Nuclear Targets}
\begin{figure}[htbp]
\begin{center}
\includegraphics*[width=9.5cm,height=13.0cm]{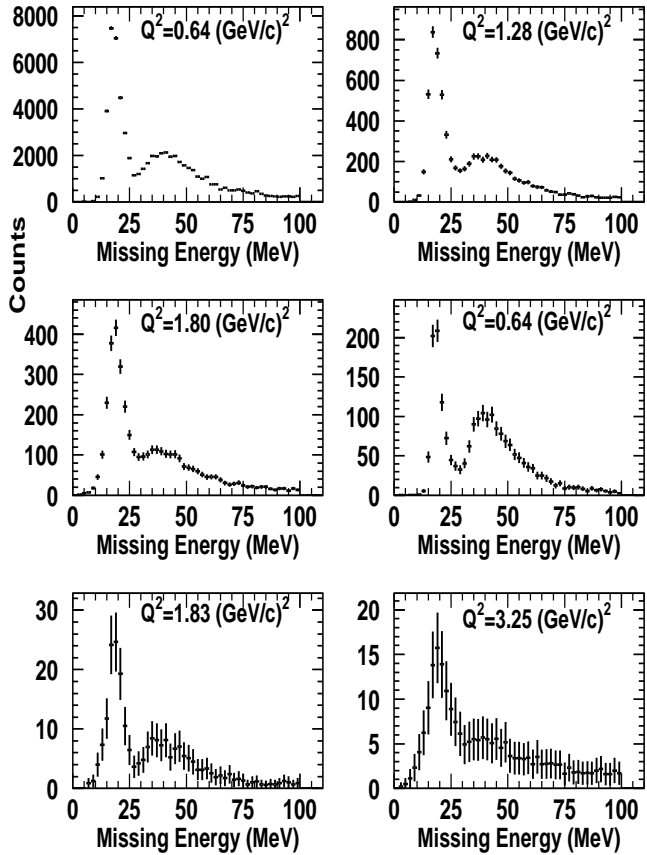}
\end{center}
\caption{Measured missing energy spectra for carbon at the different Q$^2$, panels with the same Q$^2$ refer to the forward and backward electron angle kinematics respectively for the L/T separation kinematics. }
\label{carbonem}
\end{figure}

\begin{figure}[htbp]
\begin{center}
\includegraphics*[width=9.5cm,height=13.0cm]{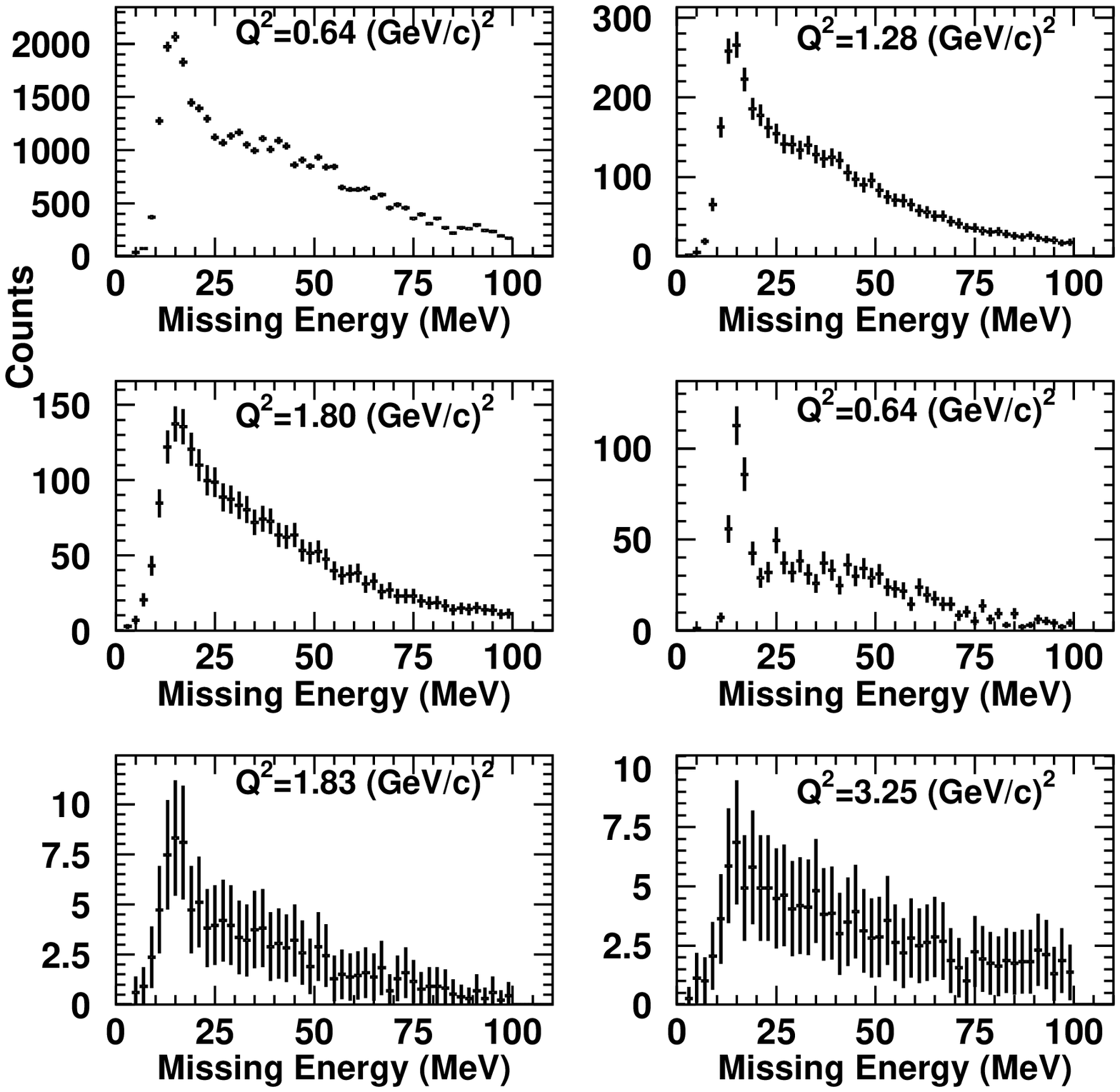}
\end{center}
\caption{Measured missing energy spectra for iron at the different Q$^2$, panels with the same Q$^2$ refer to the forward and backward electron angle kinematics respectively for the L/T separation kinematics.}
\label{ironem}
\end{figure}

\begin{figure}[htbp]
\begin{center}
\includegraphics*[width=9.5cm,height=13.0cm]{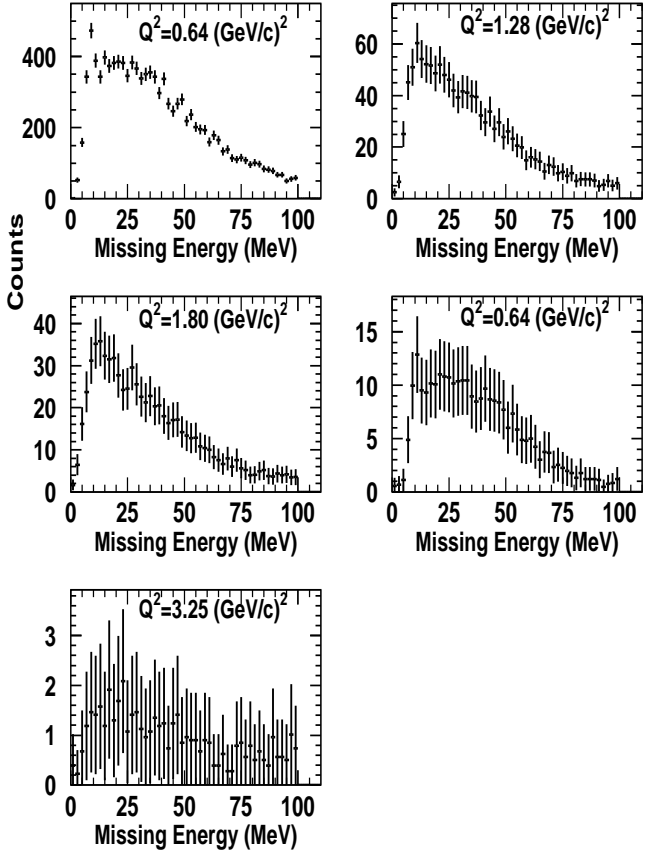}
\end{center}
\caption{Measured missing energy spectra for gold at the different Q$^2$, panels with the same Q$^2$ refer to the forward and backward electron angle kinematics respectively for the L/T separation kinematics.}
\label{goldem}
\end{figure}

A missing energy and missing momentum spectrum was obtained at each data point 
for all three nuclear targets. These are the raw spectra from which the spectral 
functions were extracted after unfolding the radiative effects, the phase space 
weight and the $e-p$ cross-section weight. The raw 
missing energy spectra are shown in Figs.~\ref{carbonem},~\ref{ironem}, and 
~\ref{goldem}. 

Fig.~\ref{carbonem} shows the missing energy spectra for carbon.  At Q$^2$ = 
0.64 (GeV/c)$^2$ the spectra show a rather sharp peak corresponding to 
populating low-lying levels in $^{11}$B which can be attributed to removing p~-~shell protons from $^{12}$C and a broader peaking at higher missing energies 
which is primarily due to removing s~-~shell protons.  The valley between the two 
groups is increasingly filled in as Q$^2$ increases, because the (absolute) 
energy resolution broadens as the energy of the particles increases, as noted 
above in discussing the hydrogen spectra of Fig.~2.  At the 
two values of Q$^2$ at which L~-~T separations were performed the valley between the s~-~shell and p~-~shell region is less distinct at the forward electron angle, again reflecting the poorer energy resolution that was also observed in the hydrogen spectra.  The missing energy spectra for iron are shown in Fig.~\ref{ironem}. 
The ground-state region peak is more prominent at low Q$^2$ and backward angles. 
The missing energy spectra for gold are shown in Fig.~\ref{goldem}. The 
statistical uncertainties are much poorer for gold than for the other targets 
and no trends are apparent.

\subsubsection*{Radiative and Acceptance Corrections}
As previously noted, energy and momentum are lost by the electrons radiating 
photons in the Coulomb field of the target nucleus both before and after the 
scattering.  The electrons can also emit bremsstrahlung radiation in passing 
through material in the spectrometers.  The net result is a distortion of the 
spectra and the corrections to this distortion are model dependent.  The code 
SIMC was used to generate correction factors for "deradiating" the observed 
spectral functions.  Model spectral functions were used to populate bins 
in $p_m$ and $E_m$ space with both the radiative corrections turned on and 
turned off and the ratio was applied as a correction factor, bin by bin, to 
the spectral functions derived from the experimental data.  The Monte Carlo 
was also used to calculate the experimental phase space for each ($E_m$,$p_m$) bin. The experimental counts in 
each ($E_m$,$p_m$) bin corrected for radiation and divided by the phase space 
for that bin was used to obtain the "experimental" spectral function:
$$
S^{\mbox{derad}}(E_{m},p_{m})~~~=~~~~~~~~~~~~~~~~~~~~~~~~~~~~~~~~~~~~~~~~~~~~  
$$
\begin{equation}
 \frac{1}{{\cal L}~H(E_m,p_m)}\sum_{\mbox{counts}}{\frac{1}{\sigma_{ep}E_{e'}p_{p'}(E_m,p_m)}}C^{{\mbox{rad}}}(E_m,p_m)
\end{equation}

\noindent
where ${\cal L}$ is the luminosity, $H(E_m,p_m)$ the phase space for the 
given $E_m$, $p_m$ bin, $C^{{\mbox{rad}}}(E_m,p_m)$ the correction factor for 
the same bin and $\sigma_{ep}E_{e'}p_{p'}(E_m,p_m)$ the off-shell $e-p$ cross-
section and kinematic factors averaged over the $E_m$ and $p_m$ bin.  This  
"experimental" spectral function is then compared to the input model spectral 
function and if the two differ by more than a specified amount the 
experimental spectral functions become the new model spectral functions and 
the whole process is iterated until a satisfactory convergence is achieved.  
In order to test the validity of this procedure non-physical spectral 
functions were input as the model spectral functions and it was demonstrated 
that after several iterations the extracted spectral functions are virtually 
independent of the initial model function.  The consistency of this 
de-radiation procedure was also checked using Monte Carlo generated data. It should be noted that these 
corrected spectral functions still include distortions due the effects of 
final state nuclear interactions, including absorption.  

\subsubsection*{Experimental Spectral Functions}

\begin{figure}[htbp]
\begin{center}
\includegraphics*[width=9.5cm,height=12.0cm]{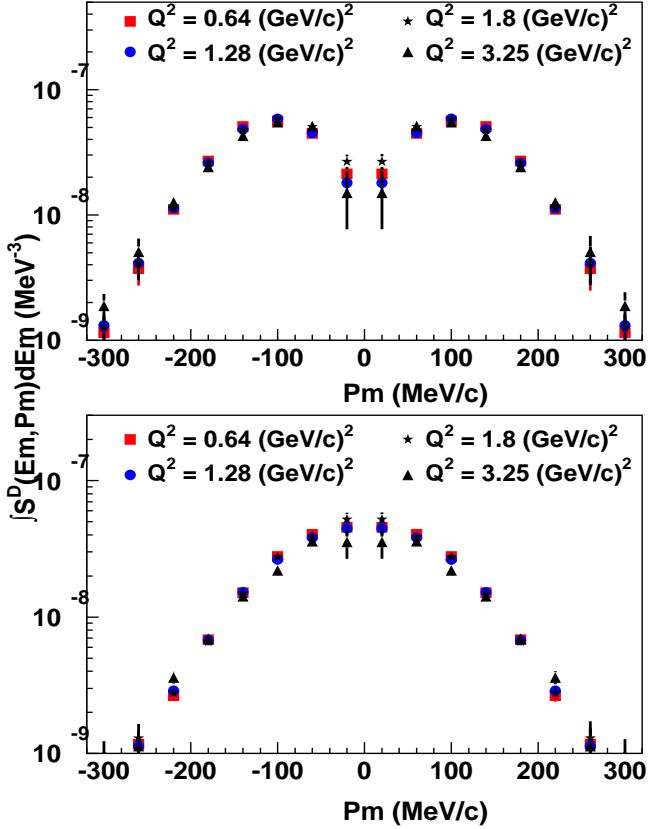}
\end{center}
\caption{Momentum distributions for carbon p~-~shell (top panel, 10$<$$E_m$$<$25~MeV) and s~-~shell (bottom panel, 30$<$$E_m$$<$50~MeV). They have been normalized so that the integral of the measured 
spectral functions over $|p_{m}|<$~300 MeV/c is equal to the integral of the 
spectral function at Q$^2$ of 1.8 (GeV/c)$^2$.}
\label{carbonpm}
\end{figure}

\begin{figure}[htbp]
\begin{center}
\includegraphics*[width=9.0cm,height=9.0cm]{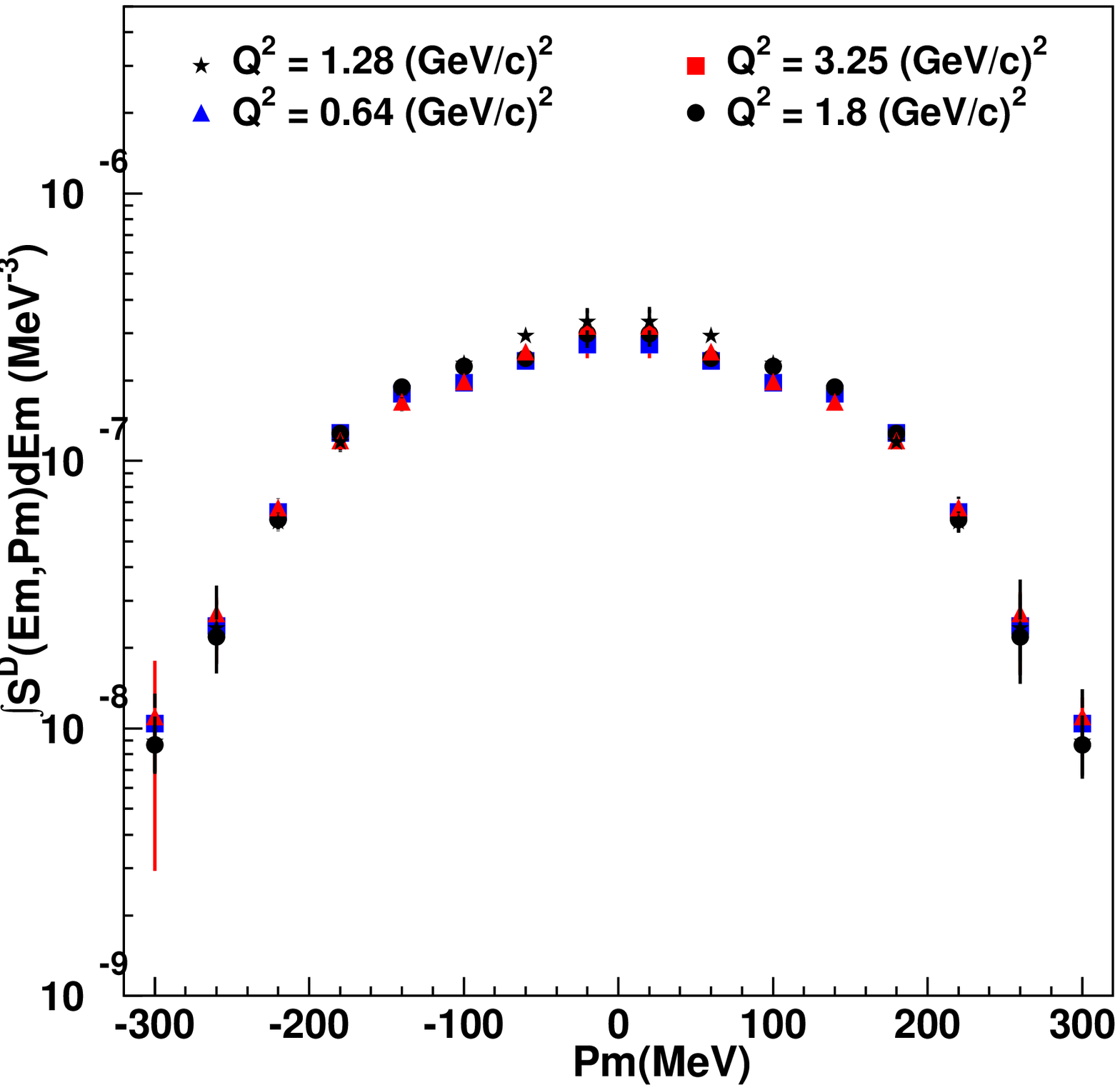}
\end{center}
\caption{Momentum distributions for iron integrated over an $E_m$ range 0$<$$E_m$$<$80~MeV. They have been normalized so that the integral of the measured spectral functions over $|p_{m}|<$~300 MeV/c is equal to 
the integral of the spectral function at Q$^2$ of 1.8 (GeV/c)$^2$.}
\label{ironpm}
\end{figure}

\begin{figure}[htbp]
\begin{center}
\includegraphics*[width=9.0cm,height=9.0cm]{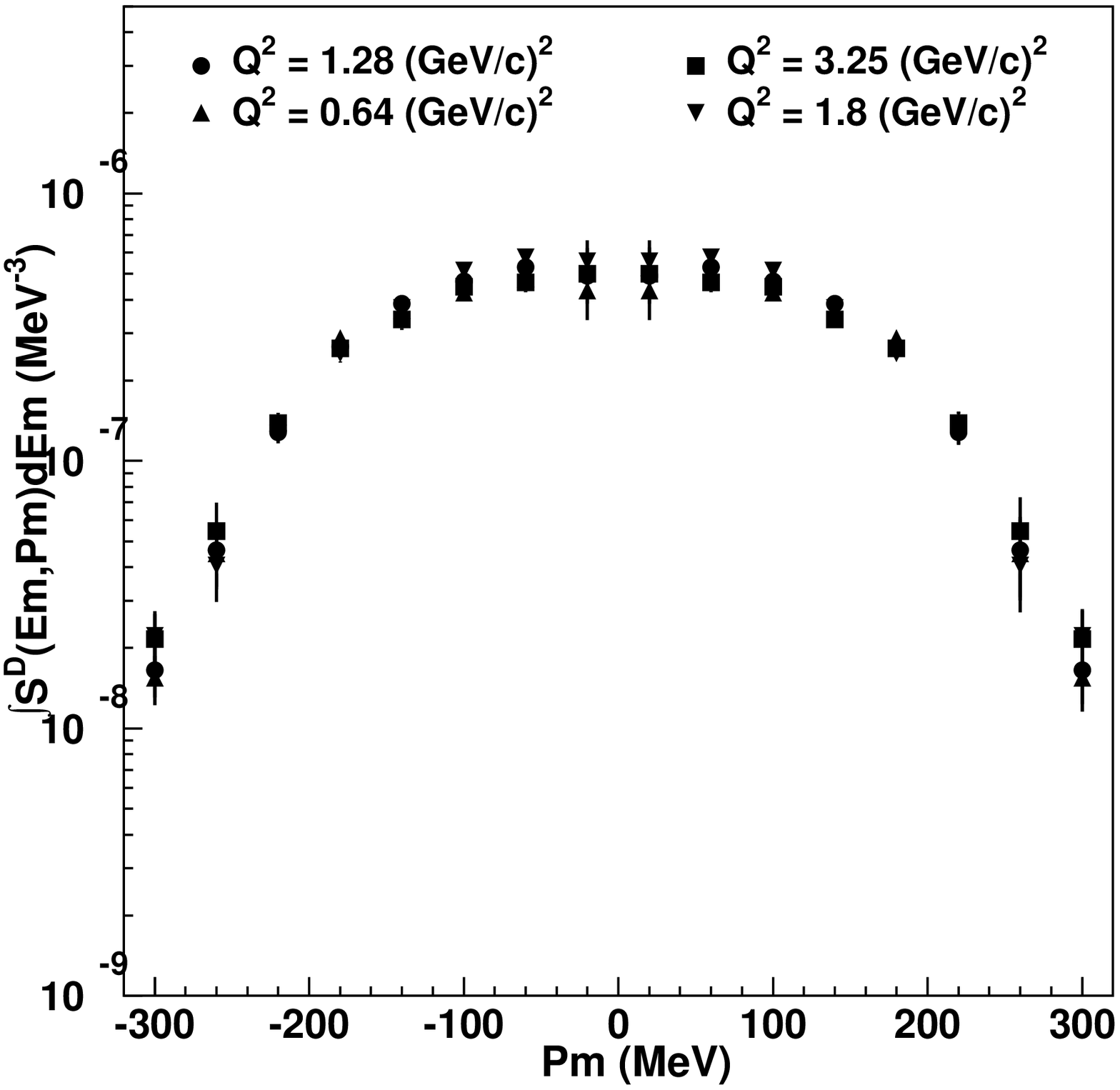}
\end{center}
\caption{Momentum distributions for gold integrated over an $E_m$ range 0$<$$E_m$$<$80~MeV. They have been normalized so that the 
integral of the measured spectral functions over $|p_{m}|<$~300 MeV/c is equal 
to the integral of the spectral function at Q$^2$ of 1.8 (GeV/c)$^2$.}
\label{goldpm}
\end{figure}

At each electron angle the above procedure was used for each proton angle to 
obtain experimental (distorted, as defined above) spectral functions and these 
were integrated over the proton angles to obtain the experimental spectral 
functions for that target, electron angle and Q$^2$.  These summed spectral 
functions are functions of both missing momentum and missing energy and 
therefore the missing momentum was integrated over in order to obtain the 
energy spectral functions and the missing energy was integrated over to 
obtain momentum distributions. The momentum distributions are 
shown in Figs.~\ref{carbonpm},~\ref{ironpm} and ~\ref{goldpm}.
 
The carbon momentum distributions are shown in Fig.~\ref{carbonpm}. They have been normalized to the spectral functions at Q$^2$ of 1.8 (GeV/c)$^2$ to remove the
effect of variation in final state interactions between the different Q$^2$ 
points.
These spectra show little variation with Q$^2$. The dip at zero 
missing momentum for missing energy between 10 and 25 MeV is attributable to 
the fact that the protons in this energy region are primarily l = 1 while only l = 0 
protons can have zero missing momentum. There is a left-right (or $\pm$) 
asymmetry in the momentum distributions that is discussed below. As with carbon 
the iron momentum distributions (Fig.~\ref{ironpm}) and gold momentum 
distributions (Fig.~\ref{goldpm}) show little change 
with Q$^2$.

\subsubsection*{Independent Particle Shell Model}

Model spectral functions were calculated in the Independent Particle Shell 
Model (IPSM) approximation, in which the nucleus is considered a sum of 
nucleons occupying distinct shells with each proton in the lowest possible 
shell. The parameters of the spectral function were adjusted to reproduce data
from low-Q$^2$ A$(e,e'p)$ and A$(p,2p)$ experiments. For $^{12}$C the removal energy 
and energy width of the two shells, s$_{1/2}$ and p$_{3/2}$ is based on the Saclay $^{12}$C$(e,e'p)$ data~\cite{moug76}. The removal energy and energy width for the $^{56}$Fe shells were based on the $^{58}$Ni$(e,e'p)$ data from Saclay~\cite{moug76, fesaclay}, with the removal energy corrected for the 2~MeV difference between $^{56}$Fe and $^{58}$Ni. The removal energy for the shells not resolved in the Saclay experiment were obtained from Hartree-Fock calculations~\cite{quintth} and the widths for these shells were calculated according to the Brown and Rho~\cite{rho81} parametrization of data for A~$<$~58. 
Similarly for $^{197}$Au the removal energies and widths are based on those 
measured for nearby nucleus $^{208}$Pb in  A$(e,e'p)$ experiments at NIKHEF~\cite{quintth}, with removal energies corrected for the 2.2~MeV difference between  $^{208}$Pb and $^{197}$Au. The parameters for the unmeasured shells were obtained 
from Hartree-Fock calculations~\cite{quintth} and the Brown and Rho parametrization as mentioned above. Further details are given elsewhere~\cite{thesis}. 

Momentum distributions were obtained for each shell by solving the 
Schr\"{o}edinger equation in a  Woods-Saxon potential using the code 
DWEEPY~\cite{guisti87}.  For  $^{12}$C the parameters used in the potential 
were based on the Saclay $^{12}$C$(e,e'p)$ data~\cite{moug76}. The $^{56}$Fe 
and  $^{197}$Au momentum distributions were based on those measured for the 
nearby nucleus $^{58}$Ni and $^{208}$Pb, modified to agree with the $^{56}$Fe$(e,e'p)$ and $^{197}$Au$(e,e'p)$ data from SLAC experiment NE-18~\cite{tonth}, respectively. For   $^{56}$Fe and  $^{197}$Au a Perey factor (with $\beta$~=~0.85)~\cite{perey63} was used to correct for the non-locality or energy dependence 
of the potential.

\begin{figure}[htbp]
\begin{center}
\includegraphics*[width=9.0cm,height=10.0cm]{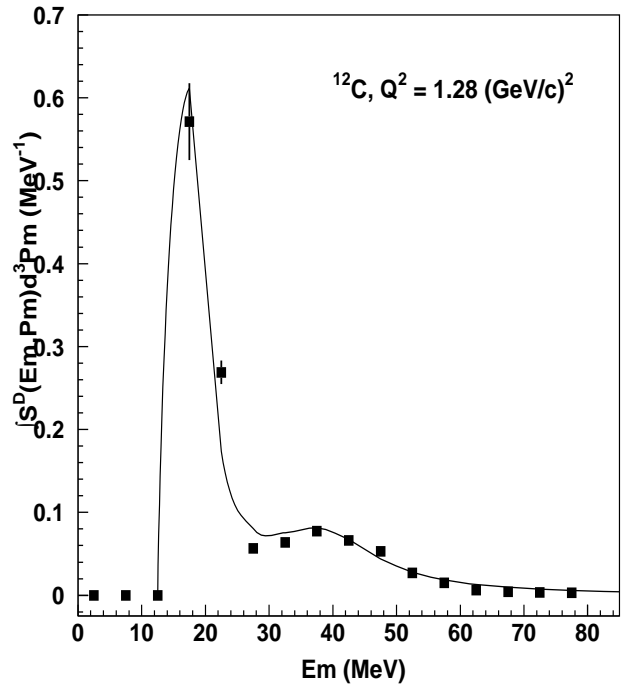}
\end{center}
\caption{Measured missing energy spectral function for carbon at Q$^2$ = 1.28 
(GeV/c)$^2$ compared to Independent Particle Shell Model (IPSM).}
\label{carbonsem}
\end{figure}

\begin{figure}[htbp]
\begin{center}
\includegraphics*[width=9.0cm,height=10.0cm]{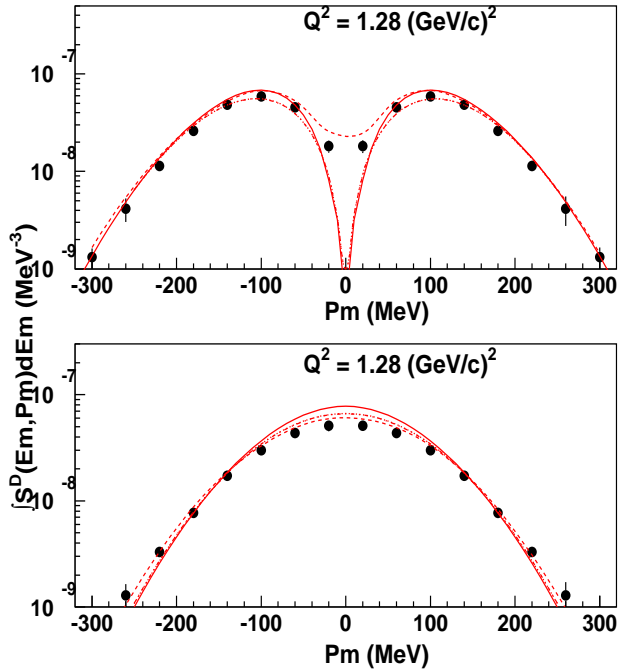}
\end{center}
\caption{Measured momentum distribution for carbon at Q$^2$ = 1.28 
(GeV/c)$^2$ in the s-removal energy region (top panel, 10$<$$E_m$$<$25~MeV) and p-removal energy region (bottom shell, 30$<$$E_m$$<$50~MeV) compared to theoretical predictions. The solid line is the 
IPSM model; dashed line is IPSM with an 8\% s-p mixing. Dotted line is a DWIA 
calculation from Zhalov {\it et al.} ~\cite{zalsf} and the dot-dashed line is 
the same DWIA calculation with color transparency included.}
\label{carbonspm}
\end{figure}

The experimental missing energy spectral function for carbon at Q$^2$ = 1.28 
(GeV/c)$^2$ is compared to the IPSM spectral function in Fig.~\ref{carbonsem}.  
The model predicts slightly too much yield in the dip region between the 
s$_{1/2}$ and the p$_{3/2}$ shells possibly implying that the s~-~shell is more 
tightly bound than generally thought.  The momentum distribution 
(Fig.~\ref{carbonspm}) in the region of the low missing energy peak, considered 
to be the p~-~shell region, shows a much shallower minimum at $p_m$ = 0 than the 
IPSM prediction, while for protons from the s~-~shell region the $p_m$ = 0 yield is 
smaller than predicted.  Agreement is much better if an 8\% p~-~s mixing is 
included (the $E_m$ cut allows some s-shell strength into the p-shell region and vice-a-versa). The spectroscopic factors found in a high-resolution (e,e'p) 
experiment done at NIKHEF~\cite{gerard88} support the amount of s~-~p "mixing" 
invoked to explain the carbon missing momentum distributions.

\begin{figure}[htbp]
\begin{center}
\includegraphics*[width=9.0cm,height=10.0cm]{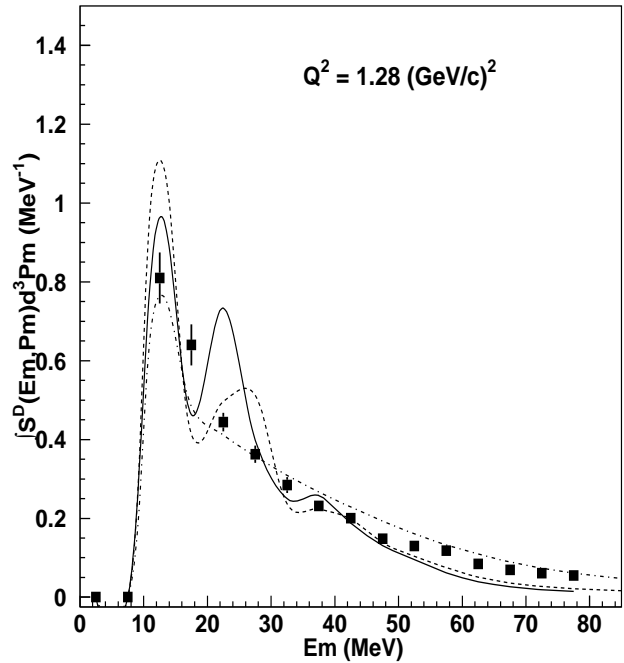}
\end{center}
\caption{Measured missing energy spectral function for iron at Q$^2$ = 1.28 
(GeV/c)$^2$ compared to theoretical models. The solid line is using the IPSM 
model.  The dashed line is a calculation from Benhar {\it et 
al.}~\cite{benharsf} and the dot-dashed line is from calculations using the 
TIMORA code described in Ref.~\cite{hor81} with spreading widths taken from 
the IPSM.}
\label{ironsem}
\end{figure}

The IPSM predicts sharper structure in the iron missing energy spectral 
functions (Fig.~\ref{ironsem}) than is observed indicating that the shell widths 
are underestimated.  This model also predicts too few of the most loosely bound 
nucleons.  Similar differences between calculation and 
experiment are seen in the gold data (Fig.~\ref{goldsem}).  For both iron and gold the momentum spectral functions are fairly well predicted although in both cases the yield 
for $|p_m|>$~250 MeV/c is under-predicted, which is probably because the 
calculations under-estimate the contribution from short-range correlations.  It must be emphasized that in 
obtaining the 
transparencies, discussed in the next section, the data were integrated out to a 
missing energy of 80 MeV and therefore differences in spectral function 
structure between model and experiment are pretty well averaged out.

\begin{figure}[htbp]
\begin{center}
\includegraphics*[width=9.0cm,height=10.0cm]{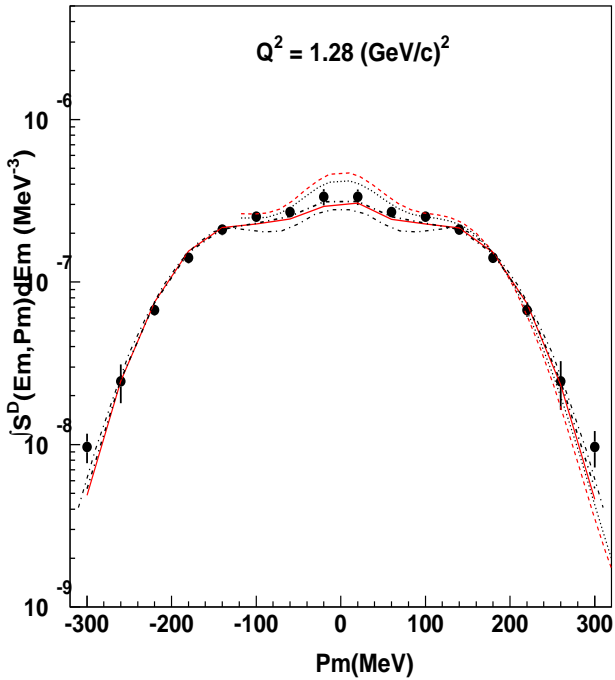}
\end{center}
\caption{Measured momentum distribution for iron integrated over an 
$E_m$ range 0$<$$E_m$$<$80~MeV at Q$^2$ = 1.28 (GeV/c)$^2$, 
compared to theoretical predictions. Solid line is using the IPSM model.  
Dotted line is DWIA calculation from Zhalov {\it et al.}~\cite{zalsf} without 
including color transparency and dot-dashed is the same with color 
transparency included.  Dot-dot-dash line is a calculation from 
Benhar {\it et al.}~\cite{benharsf} and dash-dot-dash line is from calculations 
using the TIMORA code described in Ref.~\cite{hor81}.} 
\label{ironspm}
\end{figure}

\begin{figure}[htbp]
\begin{center}
\includegraphics*[width=9.0cm,height=10.0cm]{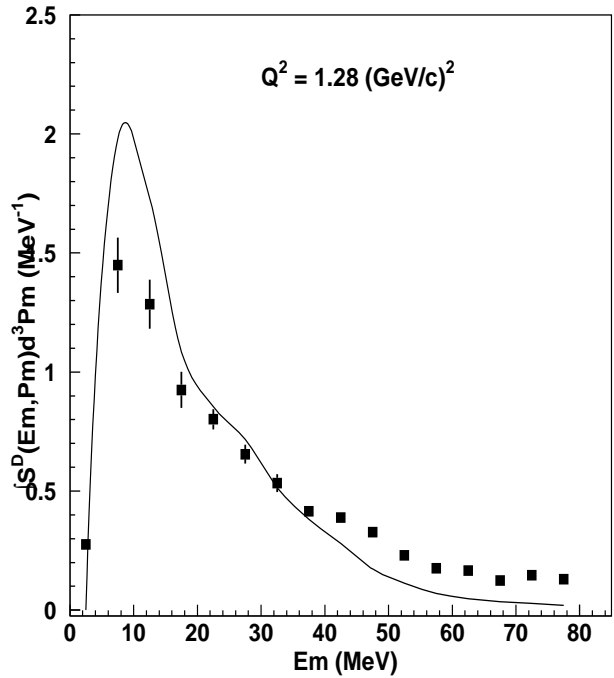}
\end{center}
\caption{Measured missing energy spectral function for gold at Q$^2$ = 1.28 
(GeV/c)$^2$ compared to the IPSM model.}
\label{goldsem}
\end{figure}

\begin{figure}[htbp]
\begin{center}
\includegraphics*[width=9.0cm,height=10.0cm]{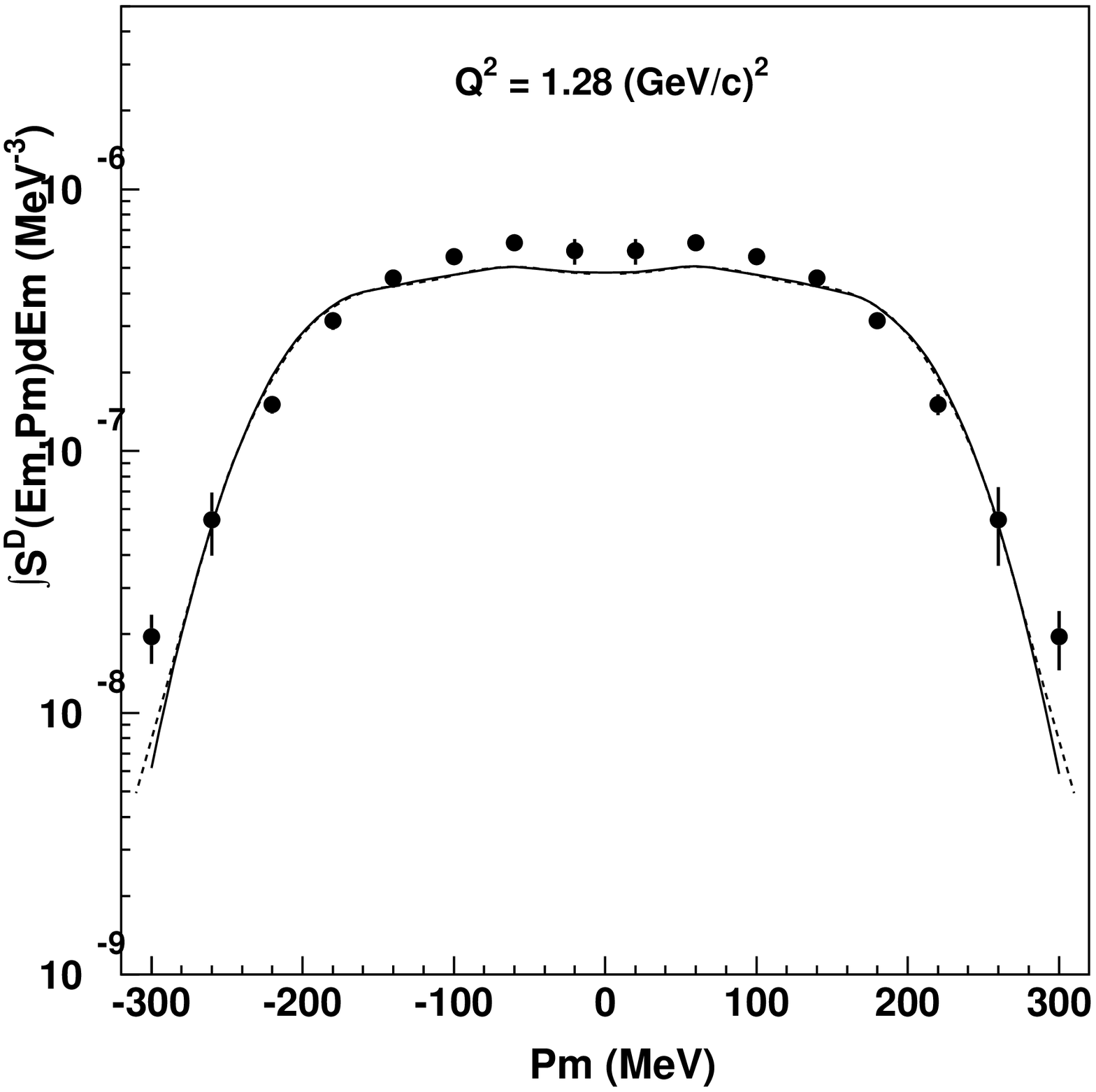}
\end{center}
\caption{Measured momentum distribution for gold integrated over an 
$E_m$ range 0$<$$E_m$$<$80~MeV at Q$^2$ = 1.28 (GeV/c)$^2$ 
compared to theoretical predictions. Solid line is using the IPSM model and dashed line is a calculation from Benhar {\it et al.}~\cite{benharsf}} 
\label{goldspm}
\end{figure}

\subsubsection*{Other Calculated Spectral Functions}

Distorted Wave Impulse Approximation (DWIA) calculations of the (distorted) 
spectral functions using the Hartree-Fock model with Skyrme's interaction to 
describe the single particle aspects of the nuclear structure~\cite{llf94} have 
been performed by Zhalov~\cite{zalsf}.  These calculations include an estimate 
of the effects of color transparency, which are negligible for carbon 
(Fig.~\ref{carbonspm}) and barely discernible in iron (Fig.~\ref{ironspm}). 
These calculations overestimate the yield at small missing momentum and fall off 
too rapidly at large $|p_m|$. 
Spectral functions have also been calculated by Benhar~\cite{benharsf}.  Here 
single-particle spectral functions are modified by adding terms dependent on 
the nuclear density.  Results are shown in Fig.~\ref{ironspm} (iron) and 
\ref{goldspm} (gold).  Including the density dependence does increase the large 
$p_m$ tail, though not
by enough to reproduce the data.  These calculations also underestimate the 
$p_m$ = 0 region (it must be remembered that the momentum distribution is 
weighted by $p_m^2$ in normalizing calculation to experiment).  The calculated 
energy spectral function for iron shows more structure than is observed, 
reflecting the fact that the IPSM spreading width was also used in the Benhar 
calculation (Fig.~\ref{ironsem}). 

Energy and momentum distributions for iron have been calculated using the TIMORA 
code written by Horowitz~\cite{hor81} and based on the $\sigma - \omega$ mean 
field theory of Walecka~\cite{wel74}.  Details of this calculation are given 
elsewhere~\cite{derekth}.  As can be seen in Fig.~\ref{ironsem} this calculation 
gives a better fit to the observed structure, or lack thereof, than does either 
the IPSM or the Benhar~\cite{benharsf} calculations.

\subsection*{Transparencies}

As noted in the Introduction the basic strategy used to obtain nuclear 
transparencies was to compare the measured yield to that calculated under the 
assumption that the struck proton escapes the nucleus without further 
interaction, i.e. the transparency is defined as the ratio of the measured yield 
to that calculated using the Plane Wave Impulse Approximation, or PWIA.

\subsubsection*{PWIA}

For each target, incident electron energy, outgoing electron angle and outgoing 
proton angle, the transparency was determined as the ratio of the observed $e-p$ 
coincidence yield, integrated over missing momentum ($\pm$ 300 MeV/c) and missing 
energy (up to 80 MeV), to that calculated using the PWIA.     However, before 
the expected coincidence $e-p$ spectra in the absence of final state interactions 
can be calculated, a number of complications must be dealt with.  As its name 
implies the PWIA treats the incoming and outgoing particles as plane waves.  
There are, of course, the radiative corrections that are discussed above.  
Additionally, the incident and outgoing waves are distorted by the Coulomb field 
of the target and residual nucleus, respectively.  It has been 
shown~\cite{hor81} that these distortions can be approximated by attaching a 
phase factor to the plane wave expansion.  The acceleration by the Coulomb field 
increases the electron momentum $k$ by:
\begin{equation}
\delta k~=~f~\frac{Z~\alpha}{R}
\end{equation}  
where factor $f$ varies between 1.1 and 1.5 depending on the size of the nucleus 
and $R$ is the coulomb radius of the nucleus. This can be used to estimate the
effect of coulomb distortion on the cross-section with satisfactory 
accuracy~\cite{knoll74}.  This coulomb acceleration of the electron 
necessitates using an effective momentum 
transfer and also alters the missing momentum~\cite{ent02}.  All of these 
effects were incorporated into the PWIA and spectral functions calculations.

The PWIA calculations were done using the ``traditional'' $e-p$ free cross 
sections in which $\mu_{p}G_E/G_M$ = 1, with the ramifications of recent polarization transfer results~\cite{mjones00} discussed below in L - T separations section.  The fact that the target proton is moving and 
is bound to a nucleus (i.e. is ``off shell'') introduces considerable 
complications. Off-shell prescriptions for quasi-free $e-p$ cross sections have 
been given by deForest~\cite{defor83} and the prescription $\sigma_{cc1}$ was 
used in the present work in calculating the PWIA cross sections.  Another 
complication is the fact that the response function is no longer the incoherent 
sum of the longitudinal and transverse response functions but there are also the 
interference terms W$_{LT}$ and W$_{TT}$~(Eq.~2).  The response function 
W$_{LT}$ is anti symmetric about the conjugate, or free $e-p$ scattering, angle 
and thus vanishes in this direction, 
known as "parallel kinematics".  Of course parallel kinematics is the only 
kinematics in free $e-p$ scattering and the cross section is given by the familiar 
Rosenbluth formula. 

While it is a reasonable first approximation to take complex nuclei as a 
collection of $A$ nucleons moving in an average potential with orbits filled in 
order of increasing energy this is too simplistic a picture to use in extracting 
transparencies.  Short-range nucleon-nucleon correlations are present and one 
effect of these is to extend some single particle strength up to hundreds of MeV 
in E$_{m}$ and well beyond the Fermi momentum in p$_{m}$.  The missing 
energy spectra are indeed above the IPSM predictions at the high energy end but 
because of the acceptance cutoff of the spectrometers only a small portion of 
this "pushed-up" strength could be detected.  Under the assumption that the 
correlations produce a uniform suppression of the spectral function below the 
Fermi momentum and the missing energy limit, correlation factors of 
1.11 $\pm$ 0.03, 1.26 
$\pm$ 0.08 and 1.32 $\pm$ 0.08 for carbon, iron and gold, respectively, are 
calculated~\cite{vppw92} and these corrections have been applied to the PWIA 
cross sections in extracting the transparencies.  

\subsubsection*{Extracted Transparencies}

\begin{figure}[htbp]
\begin{center}
\includegraphics*[width=11.0cm,height=12.0cm]{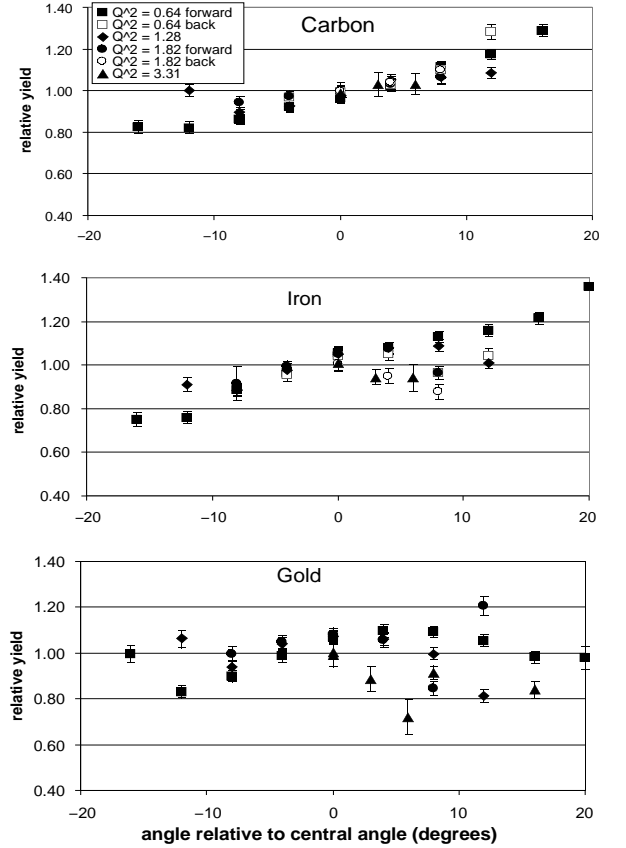}
\end{center}
\caption{Normalized transparency as a function of angle relative to the 
conjugate angle for carbon~(top), iron~(middle) and gold~(bottom). Normalization was done at the conjugate angle.} 
\label{cral}
\end{figure}

The apparent transparencies (i.e. ratio of measured to PWIA calculated (e,e'p) 
coincidence yield )  relative to that at the conjugate angle are shown in 
Fig.~\ref{cral} for the carbon~(top), iron~(middle) and gold~(bottom) targets, 
for the various electron kinematic settings. The transparencies are 
significantly asymmetric.  One possible reason could be the presence of 
interference terms in the response function, i. e. a W$_{LT}$ (Eq.~4) in excess 
of that included in the de Forest prescription $\sigma_{cc1}$.  This is not 
unexpected because modern relativistic models predict such asymmetries \cite 
{Kelley2, Udias2}. However, it should be noted that coulomb distortion of the 
electron waves can alter the effective scattering angle and therefore induce an 
asymmetry about the free conjugate angle.  While much of the coulomb distortion 
can be allowed for by introducing the momentum increase given by Eq.~10 it could 
well be that this correction is not adequate.  Coulomb distortions are known 
to increase with Z~\cite{couldistth}.  The 
angular dependence of the quasi-free scattering depends directly on the momentum distribution of the scattering nucleons and the tendency of the 
transparency to peak at the conjugate angle that is seen in the iron and gold 
distributions could be due to an underestimate of the number of high-momentum 
protons in the nucleus.  None of these complications appear to be present in the 
carbon data and so we can conclude that in carbon at least there is evidence of 
an interference term in the response function that decreases with increasing 
Q$^2$.

\begin{figure}[htbp]
\begin{center}
\includegraphics*[width=9.0cm,height=11.0cm]{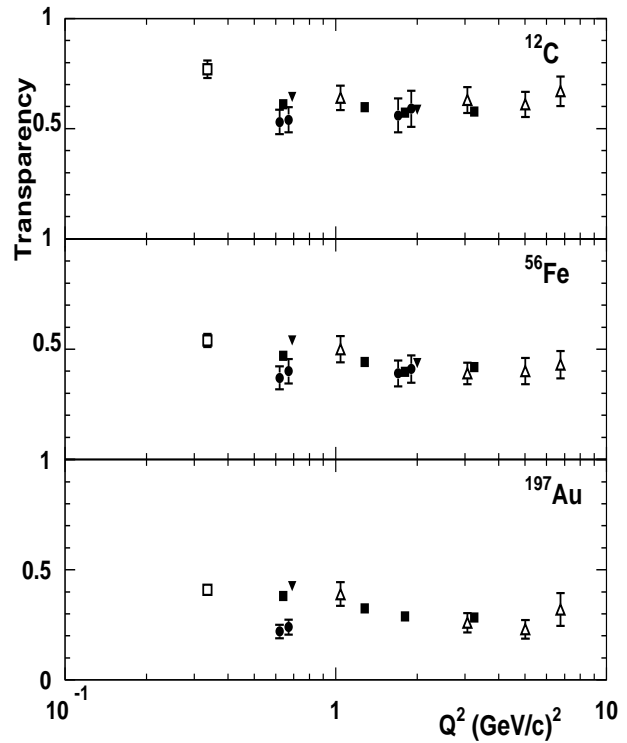}
\end{center}
\caption{Transparencies as a function of Q$^2$.  The solid squares and triangles are from the 
present work and at both 0.64 and 1.8 (GeV/c)$^2$ one of the points is slightly 
offset so that the forward and  backward angles (solid triangles) results can be 
shown separately.  Also shown are results reported from experiments at 
Bates~\cite{gerry92}(open square) and SLAC~\cite{ton95,mak94}(open triangle). The solid circles show the 
transparencies extracted from the longitudinal spectral functions extrapolated 
to all $p_m$, these have been slightly displaced in Q$^2$ for clarity.} 
\label{transp}
\end{figure}

The outgoing proton cone was integrated over in order to determine the 
transparency for that electron kinematic setting.  The values thus obtained are 
shown in Table~III and are plotted as a function of Q$^2$ for the various targets 
along with previous measurements in Fig.~\ref{transp}.  There are three types of 
errors in the transparencies:\\
(i)~Statistical:  These are down in the 0.01 region and are never greater than 
0.02.\\
(ii)~Systematic:  These are about 2.5\% overall and about 2\% from point to 
point.\\
(iii)~Model dependence:  These include uncertainties in the radiative 
corrections, the off-shell $e-p$ cross sections and the correlation corrections.  
The sum in quadrature of the model dependent uncertainties is about 5\% for C 
and 8\% for Fe and Au.  The relative uncertainties in comparing different points 
with the same target are less than 5\%.  

In addition to the obvious trend of decreasing with increasing A, the 
transparencies also decrease with increasing Q$^2$, at least at the low end of 
the Q$^2$ range covered here.  The A and Q$^2$ dependence of the transparencies 
has already been described and discussed~\cite{prl1}.  At the two values of 
Q$^2$ where data were taken at 2 different angles the transparency, as defined 
as the ratio of observed cross section to that predicted by the PWIA, is higher 
at the backward (i.e., high $\epsilon$) angle.  This is a manifestation of the 
enhancement of the transverse component of the cross section, discussed below in 
the section on the L~-~T separated spectral functions. Also shown in 
Fig.~\ref{transp} are the 
transparencies extracted from the longitudinal part of the spectral functions 
(extrapolated to include all $p_{m}$). These transparencies are lower than the 
transparencies extracted by comparing to PWIA yields and the difference 
increases with A.
  
\begin{table}
\caption{Transparencies found at the various Q$^2$ and $\epsilon$ for the 3 
targets.  Numbers in parenthesis are statistical errors only.} 
\label{transptable}
\begin{center}
\begin{tabular}{|c|c|c|c|} \hline \hline
Q$^2$  (GeV/c)$^2$ & carbon  & iron  &  gold  \\ \hline \hline
 0.64 ($\theta_e$ forward)& 0.61(0.02)& 0.47(0.01)& 0.38(0.01) \\
 0.64 ($\theta_e$ backward)& 0.64(0.02)&0.54(0.01)& 0.43(0.01) \\  
 1.28 & 0.60(0.02) &  0.44(0.01) & 0.32(0.01) \\   
 1.80 ($\theta_e$ forward)& 0.57(0.01) & 0.40(0.01)&0.29(0.01) \\
 1.83 ($\theta_e$ backward)&0.59(0.01)& 0.44(0.01) &       \\
 3.25 & 0.58(0.02) & 0.42(0.01) & 0.28(0.01)  \\ \hline \hline
\end{tabular}
\end{center}
\end{table}
  
\subsubsection*{L~-~T Separations}

L~-~T separations were performed at 0.64 and 1.8 (GeV/c)$^2$.  While at the low 
Q$^2$, small angle, point the entire cone of outgoing protons was covered just 
about as quickly as the spectrometer could be moved, because of the kinematic 
factors some compromises had to be made at the other settings. Performing L~-~T 
separations requires accurate data, partially because the anomalous proton 
magnetic moment leads to the response function being primarily transverse which, 
in turn, means that it is necessary to separate out a longitudinal response from 
a response function that is dominated by the transverse over the entire range.  
As noted above, except at the large $\epsilon$, small Q$^2$ point it was not 
possible to cover the entire cone, which would have made it possible to average 
over the interference terms in the response function.  The fact that the 
differential cross sections are not symmetric about the conjugate angle (Fig. 
15) demonstrates that these terms are not necessarily negligible.  For the L~-~T separations it was therefore decided to use only 
data where these terms must be small, namely, requiring that $|p_m|$ be less 
than 80 MeV/c. 

The spectral functions obtained using the PWIA are the weighted average of what 
can be called separated spectral functions, S$_L$ and S$_T$, and can be written:
\begin{equation}
S(E_m,{\bf p}_m) = \frac{\sigma_L~S_L~(E_m,{\bf p}_m) + \sigma_T~S_T(E_m,{\bf 
p}_m)}{\sigma_L + \sigma_T},
\end{equation}
\noindent      
and the L~-~T separation then separates out $S_L$ and $S_T$ with the deForest 
prescription~\cite{defor83} used to modify $\sigma_L$ and $\sigma_T$ from the 
free nucleon values in order to account for the fact that the nucleons are bound 
in a nucleus. The separated spectral functions for carbon have already been 
reported~\cite{prl1}.  Separated spectral functions for iron are shown in 
Fig.~17.  Because of the increasing dominance of the magnetic scattering with 
increasing Q$^2$ (Eq.~1) the errors in $S_L$ increase with increasing 
Q$^2$ while the errors in $S_T$ decrease somewhat.  The transverse strength is 
clearly smaller at the higher Q$^2$ and, at 0.64 (GeV/c)$^2$, $S_T$ is 
clearly greater than $S_L$.  At Q$^2$ = 1.8 (GeV/c)$^2$, the errors on $S_L$ are too 
great to allow any conclusions as to whether there are (relative) changes in 
$S_L$ similar in magnitude to those found in $S_T$.  Similar results were found 
for carbon~\cite{prc2}.  

An L - T separation for gold was only done at 0.64 (GeV/c)$^2$ and the resultant 
spectral functions are shown in Fig. 18.  As with the other two targets at this 
momentum transfer there is an excess of transverse strength. 

The results described above were obtained using the proton form factors 
discussed in the {\it hydrogen data} section, 
with $\mu_p$G$_E~\approx$ G$_M$. However, the 
polarization transfer measurements which now have been extended up to 5.5 
(GeV/c)$^2$ show $\mu_p$G$_E$/G$_M$~\cite{pedrisat} continuing to decrease 
approximately linearly with Q$^2$.  These ratios disagree with the series of L - T 
separation studies of $e-p$ scattering going back over 30 years which in the
aggregate~\cite{walker94,bosted94} find $\mu_p$G$_E$/G$_M$ consistent with unity 
in this momentum transfer range (and beyond). Because the spectral functions are 
close to inversely proportional to the square of the form factors large changes 
in the form factors lead to large changes in the separated spectral functions.  
\begin{figure}[htbp]
\begin{center}
\includegraphics*[width=9.0cm,height=11.0cm]{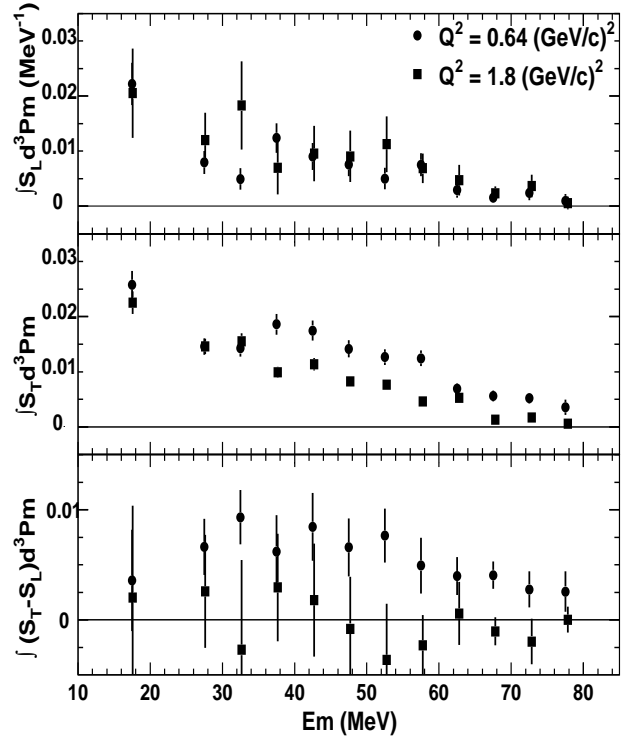}
\end{center}
\caption{Iron separated spectral functions integrated over a $p_m$ range 0$<$$p_m$$<$80 MeV/c. The Q$^2$ = 1.8 (GeV/c)$^2$ points have been displaced slightly for clarity. The lowest $E_m$ point has been averaged over 10~$<E_m<$~25~MeV. In obtaining these spectral functions the proton electric form factor was assumed to have the dipole form and the proton magnetic from factor was taken from Ref.~\cite{gk73}} 
\label{feslst}
\end{figure}

\begin{figure}[htbp]
\begin{center}
\includegraphics*[width=9.0cm,height=11.0cm]{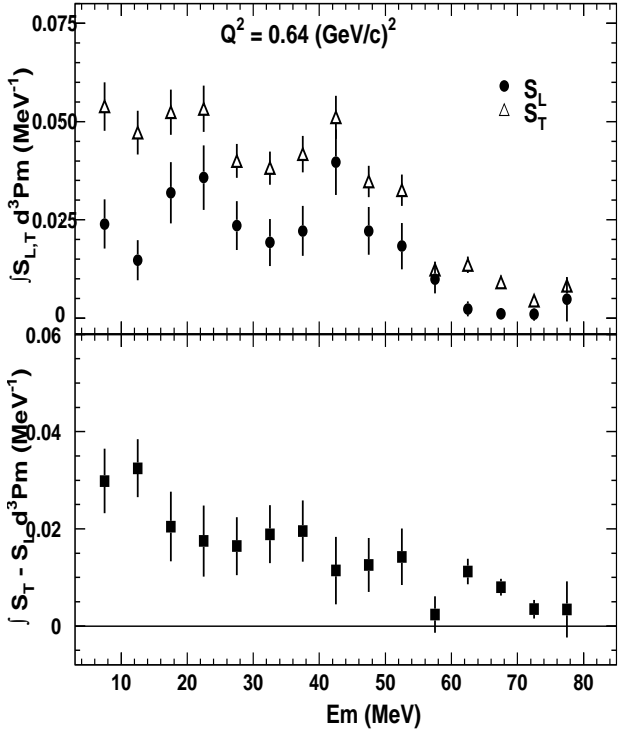}
\end{center}
\caption{Gold separated spectral functions integrated over a $p_m$ range 0$<$$p_m$$<$80 MeV/c.  In obtaining these spectral 
functions the proton electric form factor was assumed to have the dipole form 
and the proton magnetic from factor was taken from Ref.~\cite{gk73}} 
\label{auslst}
\end{figure}

\begin{figure}[htbp]
\begin{center}
\includegraphics*[width=9.0cm,height=10.0cm]{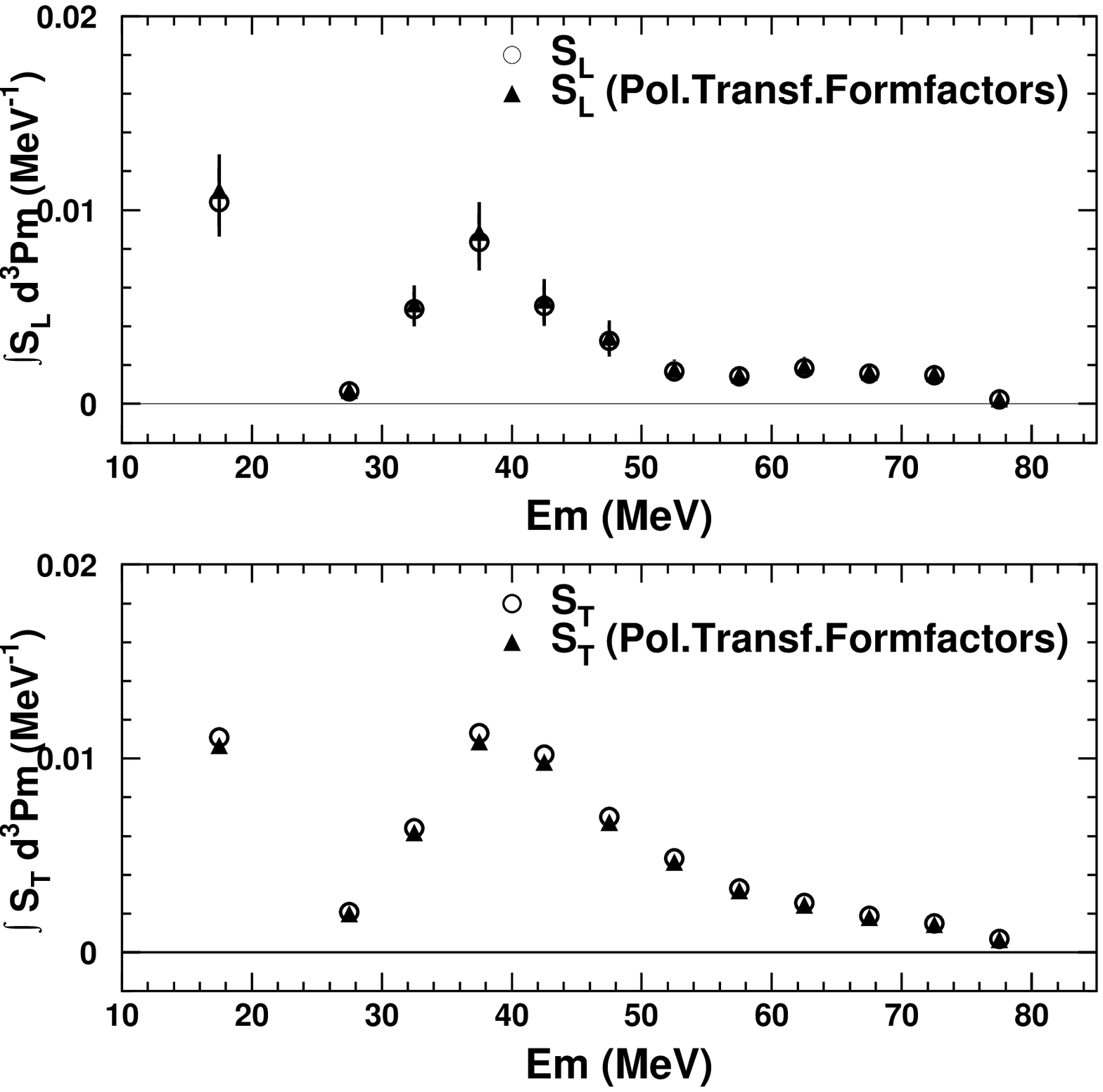}
\end{center}
\caption{Comparison of the carbon longitudinal (top panel) and transverse 
(bottom panel) spectral functions at Q$^2$ = 0.64 (GeV/c)$^2$, integrated over a
 $p_m$ range 0$<$$p_m$$<$80 MeV/c, using the proton 
form factors obtained by the Rosenbluth separation~\cite{walker94},
\cite{bosted94} (open symbols) and the polarization transfer~\cite{mjones00} 
methods (solid symbols). The lowest $E_m$ point has been averaged over 10~$<E_m<$~25~MeV.} 
\label{caslst}
\end{figure}

\begin{figure}[htbp]
\begin{center}
\includegraphics*[width=9.0cm,height=10.0cm]{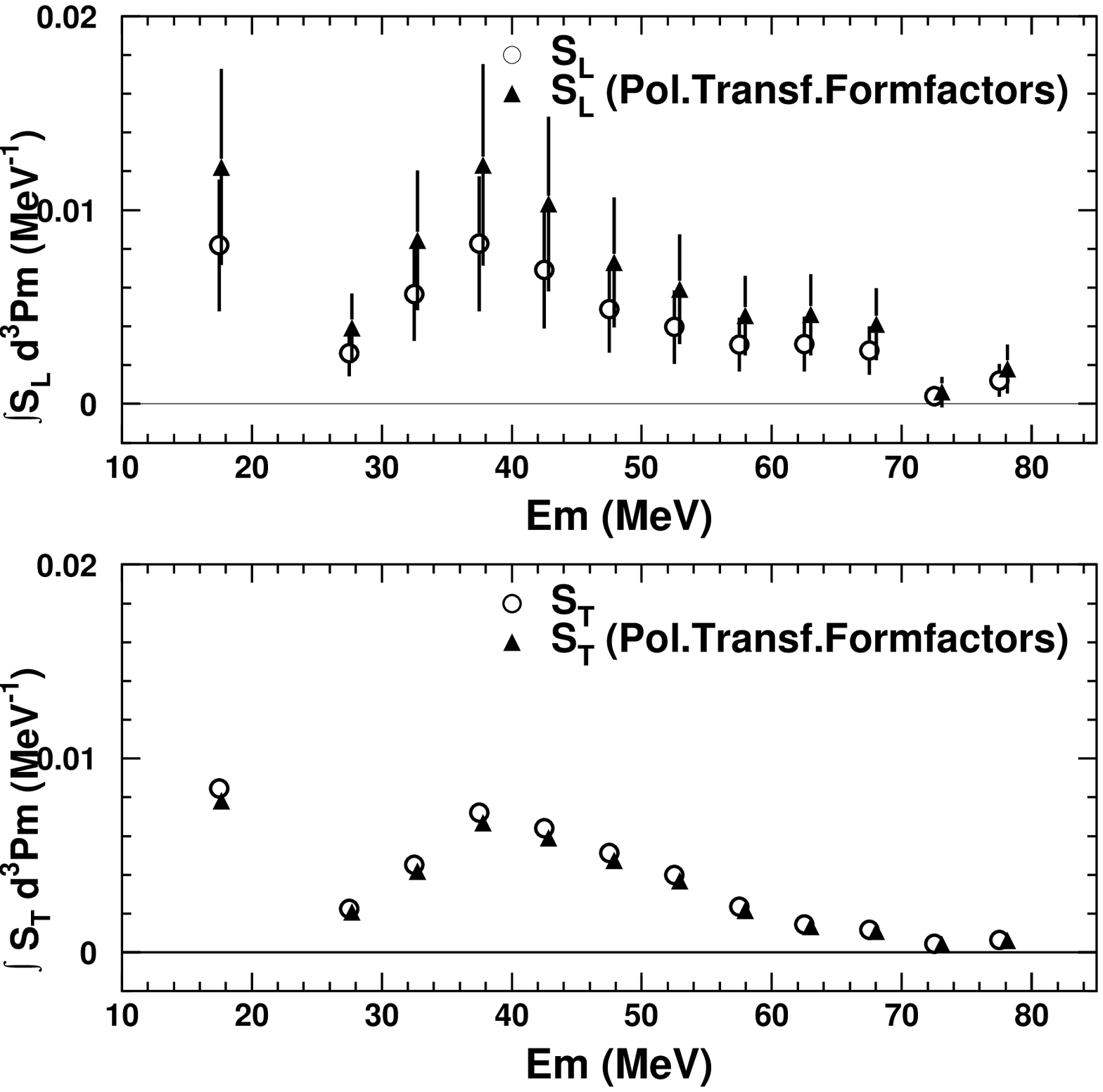}
\end{center}
\caption{Comparison of the carbon longitudinal (top panel) and transverse 
(bottom panel) spectral functions at Q$^2$ = 1.8 (GeV/c)$^2$ , integrated 
over a $p_m$ range 0$<$$p_m$$<$80 MeV/c, using the proton form factors 
obtained by the Rosenbluth separation~\cite{walker94},\cite{bosted94}  (open symbols) and the polarization transfer~\cite{mjones00} methods (solid symbols). The lowest $E_m$ point has been averaged over 10~$<E_m<$~25~MeV. The polarization transfer form factor points have been displaced slightly for 
clarity.} 
\label{ccslst}
\end{figure}

A comparison of the spectral functions obtained using the L - T 
separation~\cite{walker94},\cite{bosted94} and the polarization 
transfer~\cite{mjones00} form factors is shown in Fig. 19(20) for carbon at 
Q$^2$ =  0.64(1.8) (GeV/c)$^2$.  At 0.64 (GeV/c)$^2$ there is little effect on 
either spectral function and the decrease in transverse strength at the higher 
Q$^2$ shows little change.  However, the form factors of Ref.~\cite{mjones00} 
lead to a  60\% increase in the longitudinal strength between the two values of 
momentum transfer.  It is hard to imagine a mechanism that would lead to such a 
Q$^2$ dependency and it is clear that the final interpretation of the present (and a great deal of other) data must await a resolution of 
the question of the free proton electric form factor.

The extra transverse strength at low Q$^2$, which we attribute to multi-nucleon exchange 
currents and perhaps other multi-nucleon effects, could lead to an overestimation of the transparency because the PWIA only deals with single nucleon 
currents.  
Therefore, we also show in Fig. 16 the transparencies at Q$^2$ =  0.64 
(GeV/c)$^2$ deduced from the longitudinal spectral function alone, and these deduced transparencies are substantially lower than the nominal transparencies.  
However, we must note that the same procedure at Q$^2$ = 1.8 (GeV/c)$^2$ does not have a big effect on the deduced transparency.

The behavior of the transverse spectral function as a function of Q$^2$ is 
consistent with a recent calculation of the separated cross-sections on 
$^{16}$O~\cite{jan01}.  This calculation includes contribution 
from two-nucleon photo-absorption and predicts a reduction in the transverse 
strength with increasing Q$^2$, as observed in this experiment.  However, it 
also predicts a large effect due to the two-nucleon photo-absorption on the 
longitudinal strength which is inconsistent with the present results. It 
should be pointed out that the effects due to two-nucleon photo-absorption 
calculated in Ref.~\cite{jan01} are an upper limit rather than an exact 
prediction.

\section*{CONCLUSIONS}

Taking advantage of the high-quality electron beams and associated detection 
systems that have become available with JLab coming into operation, (e,e'p) 
coincidence measurements were made
on carbon, iron and gold targets at momentum transfers Q$^2$ of 0.64, 1.28, 1.8 
and 3.25(GeV/c)$^2$.  Spectral functions were measured for missing momentum out 
to 300 (MeV/c) and missing energy up to 80 MeV and these differ in detail, but 
not in overall shape, from Independent Particle Shell Model 
calculations.  Other reported calculations do not give much better fits except 
perhaps those from a code based on a $\sigma - \omega$ mean field theory.  By 
comparing the experimental yields integrated over missing energy and missing 
momentum with PWIA calculations nuclear transparencies for 350 - 1800 MeV 
protons were determined with an accuracy that is considerably greater than 
previously reported transparency determinations.

Longitudinal - Transverse separations were performed at 0.64 (GeV/c)$^2$ and 
1.8 (GeV/c)$^2$ with
the iron and gold separations being the first such data on medium and heavy 
nuclei.  Considerable excess transverse strength is found at Q$^2$ = 0.64 
(GeV/c)$^2$ which is much reduced at 1.8 (GeV/c)$^2$.  This excess strength is 
attributed to multi-nucleon effects that have less effect on smaller distance 
probes.  Recently reported determinations of G$_E$/G$_M$ for the proton which 
are in substantial disagreement with previously accepted values will, if they 
are confirmed, substantially alter the magnitude of the longitudinal spectral 
function at 1.8 (GeV/c)$^2$.  However, because G$_M$ is primarily determined by 
the absolute cross section the transverse spectral function will be little 
affected. 

We would like to gratefully acknowledge the outstanding efforts of the
staff of the Accelerator and Physics Divisions of Jefferson Laboratory
to make these experiments possible. This work is supported in part by the U.S. 
Department of Energy and the 
National Science Foundation.

\end{document}